\documentclass[modern]{aastex}
\usepackage{graphicx}
\usepackage{epsfig, natbib, fancyhdr}
\usepackage{amsmath}
\usepackage{epstopdf}

\newcommand{\ES}{$\Sigma_{\overline{E}}$}
\newcommand{\WS}{\textit{WISE}}
\newcommand{\WSAC}{\textit{WISE} All-Sky}

\def\hip{\textit{Hipparcos}}
\def\akari{\textit{AKARI}}
\def\iras{\textit{IRAS}}

\def\spitzer{\textit{Spitzer}}
\def\mass{\textit{2MASS}}
\def\p14{PMH14}

\accepted{ to Astronomical Journal on December 2, 2016}

\begin{document}

\shortauthors{Patel, Metchev, Heinze, Trollo}
\shorttitle{Faint \WS\ Debris Disks}

\title{The Faintest \WS\ Debris Disks:\\ Enhanced Methods for Detection and Verification}

\author{Rahul I.\ Patel\altaffilmark{1}, Stanimir A.\ Metchev\altaffilmark{2,3}, Aren Heinze\altaffilmark{4}}
\and
\author{Joseph Trollo\altaffilmark{2}}
\altaffiltext{1}{Infrared Processing and Analysis Center, California Institute of Technology, Pasadena, CA 91125}
\altaffiltext{2}{Department of Physics \& Astronomy, Centre for Planetary Science and Exploration, The University of Western Ontario, 1151 Richmond Street, London, Ontario, N6A 3K7, Canada}
\altaffiltext{3}{Department of Physics \& Astronomy, Stony Brook University, 100 Nicolls Rd, Stony Brook, NY 11794--3800}
\altaffiltext{4}{Institute for Astronomy, 2680 Woodlawn Dr., Honolulu, HI 96822--1839}

\begin{abstract}
In an earlier study we reported nearly 100 previously unknown dusty debris disks around \hip\ main sequence stars within 75~pc by selecting stars with excesses in individual \WS\ colors. Here, we further scrutinize the \hip\ 75~pc sample to (1) gain sensitivity to previously undetected, fainter mid-IR excesses and (2) to remove spurious excesses contaminated by previously unidentified blended sources. We improve upon our previous method by adopting a more accurate measure of the confidence threshold for excess detection, and by adding an optimally-weighted color average that incorporates all shorter-wavelength \WS\ photometry, rather than using only individual \WS\ colors. The latter is equivalent to spectral energy distribution fitting, but only over \WS\ band passes. In addition, we leverage the higher resolution \WS\ images available through the \url{unWISE.me} image service to identify contaminated \WS\ excesses based on photocenter offsets among the $W3$- and $W4$-band images. Altogether, we identify 19 previously unreported candidate debris disks. Combined with the results from our earlier study, we have found a total of \textbf{107} new debris disks around 75~pc \hip\ main sequence stars using precisely calibrated \WS\ photometry. This expands the 75~pc debris disk sample by 22\% around \hip\ main-sequence stars and by 20\% overall (including non-main sequence and non-\hip\ stars).

\end{abstract}

\section{Introduction}\label{sec:intro}

	Debris disks around main sequence stars are typically discovered by their characteristic infrared (IR) excesses.  Their fluxes at $\lambda \gtrsim$ 5\micron\ are significantly higher than would be expected from stellar  photospheric emission alone.  A debris disk can be detected by fitting a photospheric model to the shorter-wavelength (visible and near-IR) photometry, and by subtracting the fitted photosphere to check for a $\gtrsim 5 \mu$m excess.  A large number of debris disk-host stars have been found this way, using data from \iras\ \citep[e.g.,][and references therein]{Moor2006, Rhee2007, Zuckerman2001}, \spitzer\ \citep[e.g.,][]{Su2006, Bryden2006, Trilling2008, Carpenter2009}, \akari\ \citep[e.g.,][]{Fujiwara2013}, and \WS\ \citep[e.g.,][]{Cruz-SaenzdeMiera2014, Vican2014}.

A limitation of this approach is the accuracy of the determination of the underlying stellar photosphere.  Flux comparisons across wide wavelength ranges---optical/near-IR for the photosphere and mid-IR for the excess---can be uncertain by several per cent.  The combination of photometric data from different surveys (e.g., Tycho--2, SDSS, \mass, \WS, \iras ) incorporates often unknown systematic uncertainties in the photometric calibration among the survey filters.  Any stellar variability between the observation epochs also adds an unknown contribution. Thus, while the systematic color uncertainties of photospheric models are generally well below a per cent, the determination of the photospheric emission in the mid-IR is uncertain by a few per cent (1 $\sigma$).  Adding to these limitations are other data systematics, most common of which can be uncertainties in the mid-IR filter profiles and the corresponding color corrections \citep[e.g.,][]{Wright2010}.  As a result a number of previous searches for WISE excesses through SED fitting have resulted in high fractions of spurious excess detections, up to 50\% \citep[see discussion in][henceforth \p14]{Patel2014}.

Notable exceptions are the surveys of \citet{Carpenter2009}, \citet{Lawler2009}, and \citet{Dodson-Robinson2011}, who demonstrate that the Infrared Spectrograph \citep[IRS;][]{Houck2004} on \spitzer\ was the most sensitive instrument ever for detecting 10--40\micron\ photometric excesses from debris disks, with nearly twice as many detections as MIPS at 24\micron.  The advantage of IRS was in the ability to locally calibrate the stellar photospheric model over a  spectral range that is close to the excess wavelengths, and in the fact that the entire 5--40\micron\ spectrum could be obtained nearly simultaneously.

With its better sensitivity than \iras, a wavelength range that---similarly to \spitzer/IRS---samples both the 3--5\micron\ stellar photosphere and potential 10--30\micron\ excesses simultaneously, and with the advantage of full-sky coverage over \spitzer, \WS\ \citep{Wright2010} presents an opportunity to find unprecedentedly faint mid-IR excesses over the entire sky.  In particular, the greatest sensitivity to faint mid-IR excesses can be obtained by analyzing the distributions of stellar colors formed from combinations of short- (3.4\micron\ and 4.5\micron; $W1$ and $W2$, respectively) and long-wavelength (12\micron\ and 22\micron; $W3$ and $W4$, respectively) \WS\ bands: e.g., $W1-W3$ or $W2-W4$.

This approach has already been applied successfully to \WS\ data. \citet{Rizzuto2012} used it to search for excesses around Sco-Cen stars based on their $W1-W3$ and $W1-W4$ colors from  the \WS\ Preliminary Release Data Release\footnote{\url{http://wise2.ipac.caltech.edu/docs/release/prelim/}}. \citet{Theissen2014} applied a similar approach to search for excesses around M dwarfs using the Sloan Digital Sky Survey Data Release 7 and the AllWISE Data Release\footnote{\url{http://wise2.ipac.caltech.edu/docs/release/allwise/}}. 

In \p14 we implemented a color-excess search on the cross-section of the entire \WS\ All-Sky Survey Data Release\footnote{\url{http://wise2.ipac.caltech.edu/docs/release/allsky/}} and the \hip\ catalog \citep{Perryman1997}, with the goal to determine the frequency of warm debris disk-host stars within 75~pc. We identified stars with infrared excesses in the $W3$ and $W4$ bands by first filtering out 15 major types of flagged contaminants, then seeking anomalously red \WS\ colors ($W1-W3, W2-W3, W1-W4$, $W2-W4$, or $W3-W4$), and finally by visually checking for contamination by background IR cirrus. We sought color excesses in all combinations of \WS\ colors independently.

This had the advantage of not excluding stars without valid measurements in some of the \WS\ bands: for example, if $W1$ was excessively saturated, a star could still be determined to have an excess based on its $W2-W4$ or $W3-W4$ color.  However, where valid measurements exist for all \WS\ bands---the majority of cases---an optimally weighted combination of colors should have lower noise and potentially deliver greater sensitivity to faint excesses. 

We implement such an optimally weighted-color excess search on the same 75~pc \hip\ sample in the present study. We further refine our threshold determination for what constitutes a \WS\ color excess: by employing an empirically-motivated functional assumption about the behavior of \WS\ photometric errors.  Finally, we implement an automated method of rejecting stars with IR photometry contaminated by nearby point-like or extended objects.

We summarize the selection of our sample of stars in Section~\ref{sec:sampledef}. In Section~\ref{sec:single_color_stuff} we describe the improved accuracy with which we set the confidence threshold when seeking \WS\ excesses, and detail our weighting scheme when employing all available \WS\ photometry to calibrate the stellar photosphere. In Section~\ref{sec:auto_rejunwise} we describe our automated method for identifying contaminated sources from their photocenter offsets between $W3$ and $W4$.  We use these techniques to confirm or reject previously discovered \WS\ excesses and to find new ones; we summarize the results in  Section~\ref{sec:results}. In Section~\ref{sec:discussion} we discuss the differences in the results between the single- and the weighted-color excesses search approaches, and find that while the latter produces higher-fidelity IR excess detections, it is likely to miss a small fraction of bona fide excesses.


\section{Sample Definition}\label{sec:sampledef}

The sample for the present study comprises the majority of the \hip\ main sequence stars selected in \p14, with the added constraint that they should have reliable \WS\ All-Sky Catalog photometry in at least $W1$, $W2$, or $W3$. Although we identify and report excesses associated with stars within 75~pc, we use a larger volume of stars out to 120~pc for the entire analysis, as this larger population better samples the random noise and the photospheric \WS\ colors discussed in Section~\ref{sec:improved_detection}. The 120~pc ``parent sample'' of stars resides in the Local Bubble \citep{Lallement2003}, and so have little line-of-sight interstellar extinction. Hence, these stars are suitable for correlating optical and infrared colors. The 75~pc ``science sample'' of stars is a subset of the parent sample, chosen to take advantage of more accurate parallaxes, and so giving a clear volume limit to our study.

 Stars were also selected if they were outside the galactic plane ($|b|>5^{\circ}$) and constrained to the $-0.17\mbox{ mag} <B_T-V_T< 1.4\mbox{ mag}$ color range. Additional details of our selection process are outlined in \p14. These include additional automated screening to ensure photometric quality, consistency, and minimal contamination. We then corrected saturated photometry in the $W1$ and $W2$ bands using relations derived in \p14. Unlike in \p14, we now add a search for weighted $W3$ or $W4$ excesses (Section~\ref{sec:single_color_stuff}). For the weighted $W3$ excess search we require valid photometry in all of $W1$, $W2$ and $W3$, while for the weighted $W4$ excess search we require valid photometry in all four bands.

\section{Single-color and Weighted-color Excesses}\label{sec:single_color_stuff}

We define as single-color excesses those that are identified in individual \WS\ colors (Section~\ref{sec:improved_detection}).  Weighted-color excesses are those that are identified from the weighted combination of \WS\ colors.  Thus, a star can have both $W2-W4$ and $W3-W4$ single-color excesses, and a $W4$ weighted-color excess.  The existence of one or more single-color excesses is generally correlated, although not necessarily, with the existence of a weighted-color excess.


   \subsection{Improved Identification of Single-color Excesses}\label{sec:improved_detection}

        We identify single-color \WS\ excesses from the significance of their color excess as defined in Equation 2 of \p14:
        
\begin{equation}\label{eq:old_sig}
\Sigma_{E[Wi-Wj]} = \frac{Wi-Wj-W_{ij}(B_T-V_T)}{\sigma_{ij}}.
\end{equation}        
        
\noindent The numerator determines the color excess $E[Wi-Wj]$ by subtracting the mean photospheric color $W_{ij}(B_T-V_T)$ from the observed $Wi-Wj$ color.   We used the calibrations of \WS\ photospheric colors of main sequence stars from \p14 \citep[see also][]{Patel2014b}.  The significance of the excess $\Sigma_{E[Wi-Wj]}$ is obtained by normalizing by the total uncertainty $\sigma_{ij}$, which is a quadrature sum of the \WS\ All-Sky Catalog photometric uncertainties, uncertainties in the saturation correction applied to bright stars, and uncertainties in the photospheric color estimation (\p14). Throughout the rest of this paper, the significance of a single-color excess is denoted with $\Sigma_E$.

    The single-color \WS\ excesses are selected by seeking stars with $\Sigma_E$ values above a pre-determined confidence level (CL) threshold: CL=98\% at $W3$ and CL=99.5\% at $W4$. The CL can be expressed in terms of the false-discovery rate (FDR): ${\rm FDR}=1-{\rm CL}$.\footnote{In \p14 we incorrectly called the FDR the false-positive rate (FPR). See Figure~4 in \citet{Wahhaj2015} for an illustration of the difference between the two terms.} 
    We denote the $\Sigma_E$ value at CL as $\Sigma_{E_{CL}}$. As in \p14, we determine the $\Sigma_{E_{CL}}$ values for the different colors from the $\Sigma_E$ distributions themselves.   Thus, the $\Sigma_{E_{CL}}$ values for our respective 98\% and 99.5\% CL thresholds in $W3$ and $W4$ correspond to where the FDR drops below 2\% for $W3$ or below 0.5\% for $W4$ excesses.

    The FDR can be determined empirically from the $\Sigma_E$ excess distributions.  To estimate the distributions of uncertainties, we assume that the effect of random errors on $\Sigma_E$ is symmetric with respect to $\Sigma_E=0$.  This would be generally true if, as is our supposition, photometric errors are symmetrically distributed around zero.  

\begin{figure}
\centering
\begin{tabular}{cc}
\includegraphics[scale=0.4]{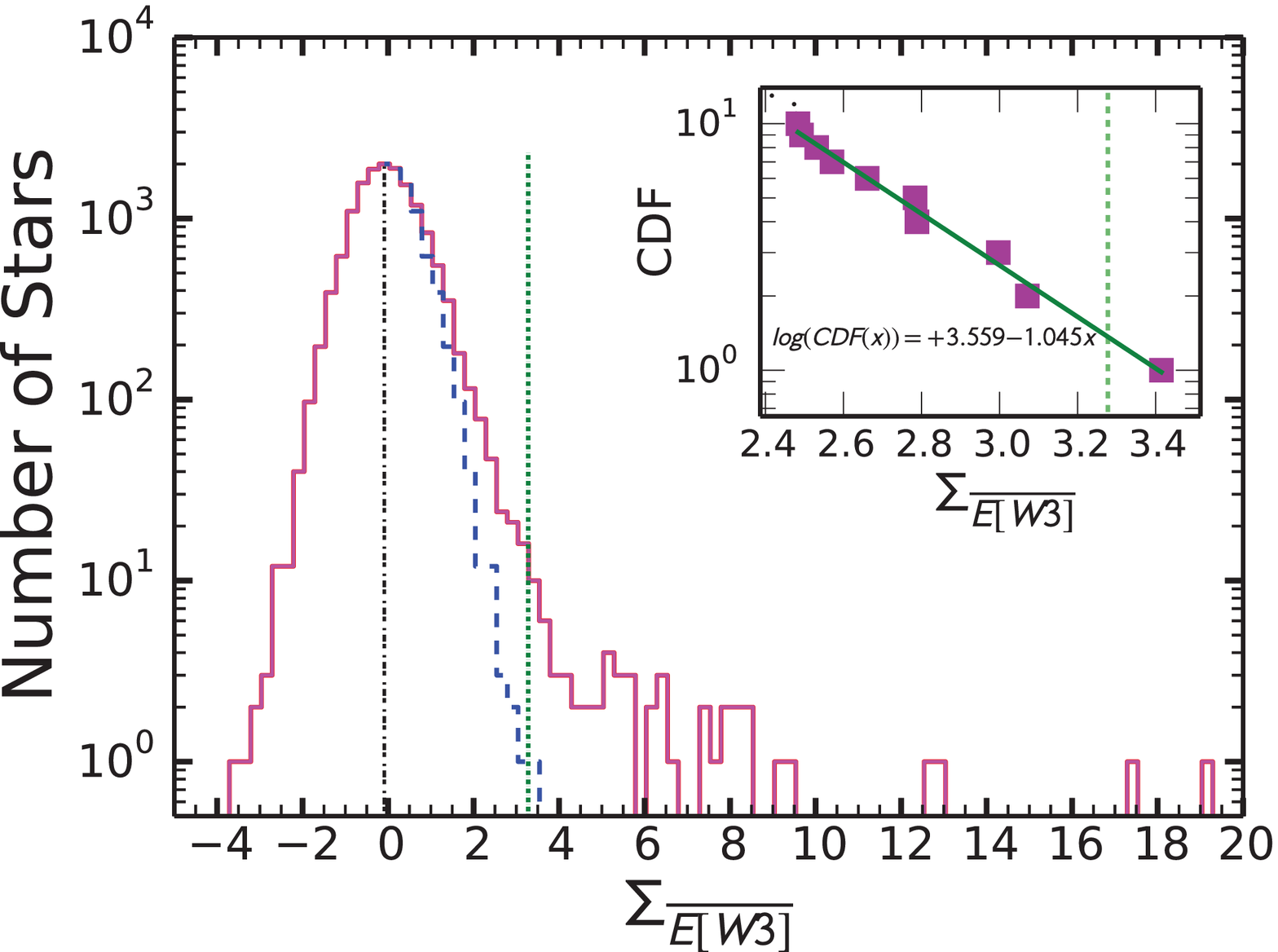}&
\includegraphics[scale=0.4]{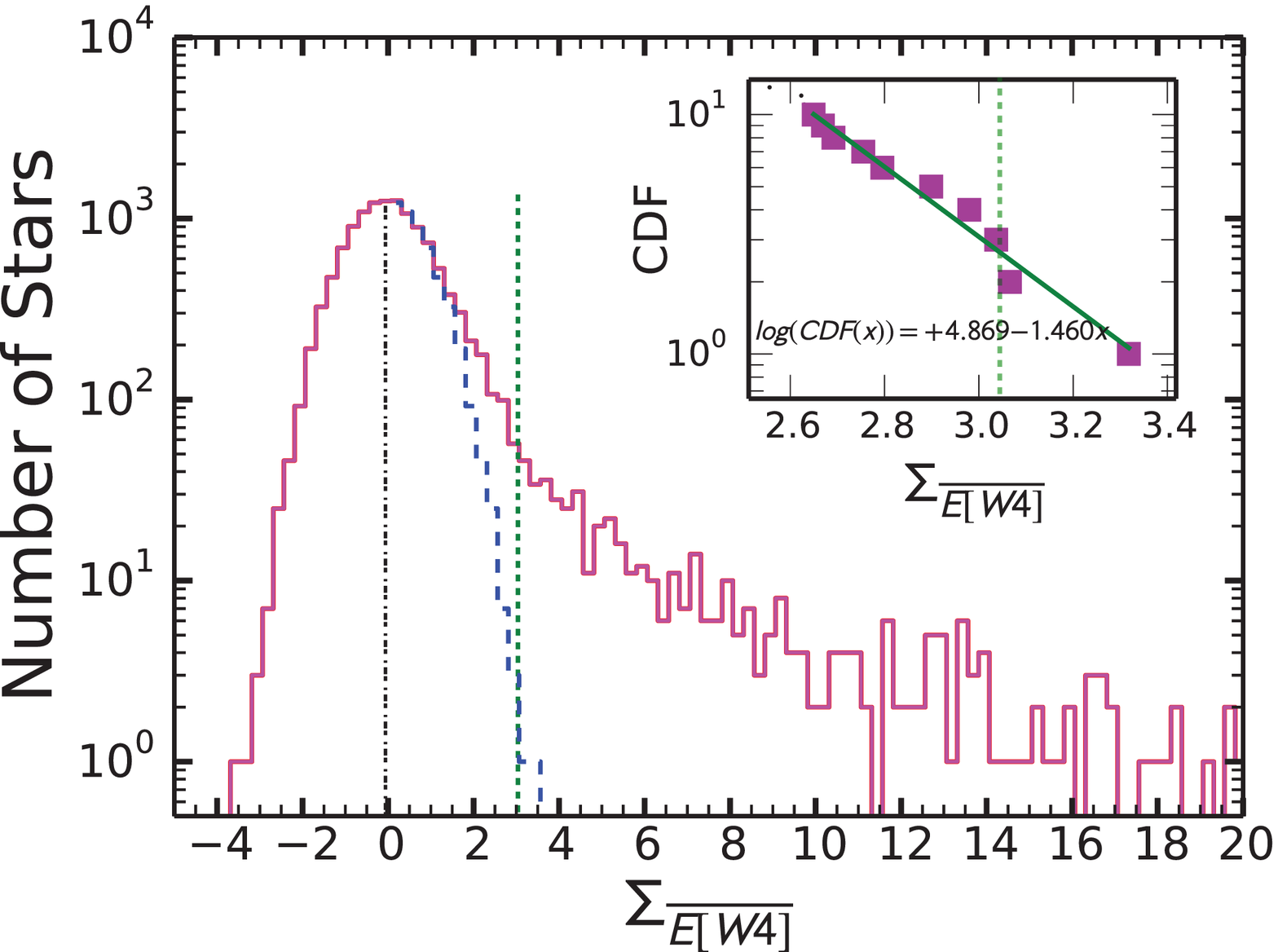}
\end{tabular}
\caption{Distributions of the weighted-color excess metrics, $\Sigma_{\overline{E[W3]}}$ (left) and $\Sigma_{\overline{E[W4]}}$ (right) for all stars in our 120~pc parent sample. We have assumed that the negative portion of each \ES\ distribution is representative of the intrinsic random and systematic noise in the data (Section~\ref{sec:improved_detection}). The mode of the full distribution is shown by a vertical black dashed-dot line. A reflection (dashed histogram) of the negative portion of the \ES\ histogram around the mode is thus representative of the false positive excess expectation. We define the FDR at a given \ES\ as the ratio of the cumulative numbers of $>$\ES\ excesses in the positive tails of the dashed and solid histograms.  The vertical dotted lines indicate the FDR thresholds for each weighted $Wj$ excess: 2\% for $W3$ and 0.5\% for $W4$.  We identify all stars with FDR values below these thresholds (correspondingly higher $\Sigma_E$ values) as candidate debris disk hosts. Each inset shows a log-log fit of a line to the last ten points in the reverse cumulative distribution function (CDF) of the uncertainties (see Section~\ref{sec:improved_detection}). Assuming exponential behavior in the tail of the uncertainty distribution, this fit smoothes over the stochasticity in this sparsely populated region of the uncertainty distribution to attain a more accurate estimate of the FDR threshold.\label{fig:ColorDist}}
\end{figure}

    The $\Sigma_E$ distributions of the various colors do indeed peak close to zero (\p14), which supports this assumption. Hence, we assume that the negative halves of the $\Sigma_E$ distributions are representative of the negative sides of the uncertainty distributions.  We then mirror the negative $\Sigma_E$ values to obtain the full distributions of uncertainties. We illustrate this method for determining the FDR in Figure~\ref{fig:ColorDist}, albeit not for the single-color excess $\Sigma_E$ metrics discussed here and in \p14, but for the weighted-color excess \ES\ metrics introduced in Section~\ref{sec:metric}.
          
    This empirical estimate of the FDR offers a straightforward method to assess the reliability of candidate excesses. However, the exact value of the $\Sigma_{E_{CL}}$ threshold tends to rely only on the one or two most-outlying stars in the (negative wing of the) $\Sigma_E$ distribution (Figure~\ref{fig:ColorDist}), and so is uncertain. In \p14 we purposefully overestimated $\Sigma_{E_{CL}}$ by the half distance to the star prior to the one that satisfied the FDR threshold. Our estimate of the $\Sigma_{E_{CL}}$ was conservative, not very accurate, and may have excluded potentially significant excesses. 
    \begin{figure}
    \centering
    \includegraphics[scale=0.4]{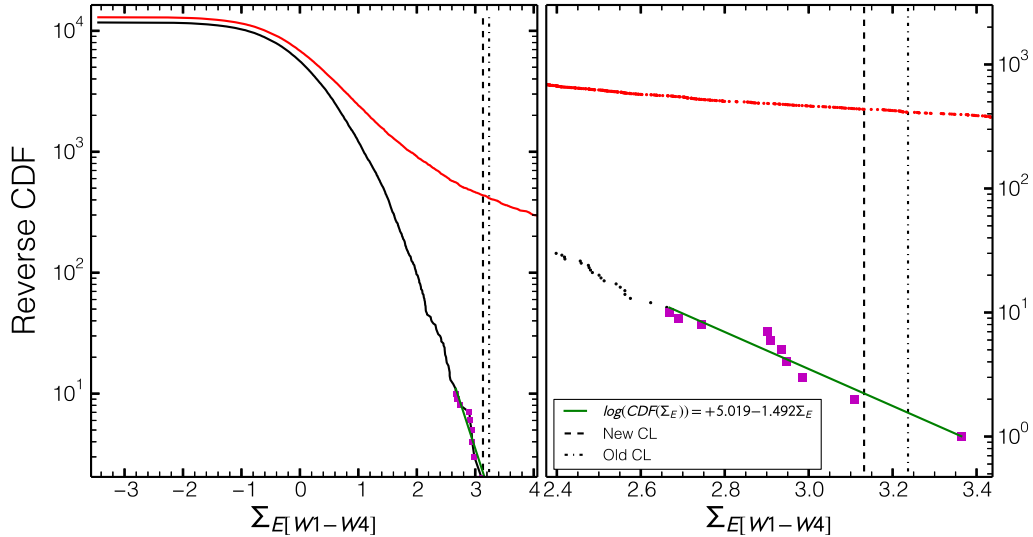}
    \caption{A reverse cumulative distribution function (rCDF, Section~\ref{sec:improved_detection}) of the uncertainty (black) and excess (red) distributions of $\Sigma_{E[W1-W4]}$. We use the rCDF to estimate the FDR at any $\Sigma_E$, with FDR being the ratio of the black and red rCDFs. The vertical dash-dotted line shows the more conservative $\Sigma_{E_{99.5}}$ estimate of the confidence threshold from \p14, set half-way between the last two points.  The vertical dashed line shows the present $\Sigma_{E_{99.5}}$ estimate, based on a fit (solid green line) to the last ten data points in the tail of the rCDF (magenta squares). The left panel shows the full rCDFs, while the right panel zooms in near the $\Sigma_{E_{CL}}$ threshold.\label{fig:CL_Comparison}} 
    \end{figure}
    
    Here we iterate on this approach by taking advantage of the near-Gaussian behavior of each uncertainty distribution. To circumvent the small-number sampling in the tail, we average the functional behavior by fitting an exponential curve to the last ten points in the reverse cumulative distribution function (rCDF) of the uncertainty distribution (Figure~\ref{fig:CL_Comparison}). This continuous form of the tail of the uncertainty distribution enables a more accurate estimate of the FDR. 

\clearpage
\begin{deluxetable}{lcccccc}
    \tablewidth{0pt}
    \tabletypesize{\scriptsize}
    \tablecaption{Single- and Weighted-Color Excess Selection Summary\label{tab:fdr_thresholds}}
    \tablehead{\colhead{Color} & \colhead{$\Sigma_{E_{\rm CL}}$\tablenotemark{a}} & \colhead{Stars in} & \colhead{Stars in} & \colhead{Excesses in} & \colhead{Debris Disk} & \colhead{New} \\
                \colhead{} &\colhead{or $\Sigma_{\overline{E_{\rm CL}}}$ } &\colhead{Parent Sample ($<$120~pc)} &\colhead{Science sample ($<$75~pc)} &\colhead{Science Sample} &\colhead{Candidates}&\colhead{Excesses}
                }
    \startdata
    $W1-W4$                 &  3.13    &  12942 & 6294 & 134 & 114 &0\\
    $W2-W4$                 &  3.06    &  13203 & 6507 & 191 & 168 &10\\
    $W3-W4$                 &  2.89    &  14434 & 7198 & 238 & 209 &12\\    
    $W1-W3$                 &  2.66    &  15017 & 6788 &  13 & 9   &1 \\ 
    $W2-W3$                 &  3.83    &  15245 & 6962 &   3 & 3   &0\\
    \hline
    Weighted $W4$           &  3.04    &  12654 & 6140 & 188 & 166 & 1\\
    Weighted $W3$           &  3.28    &  14808 & 6684 &   6 & 6   & 0\\        
    \hline\hline\\
    Total                   & \nodata  &  16960 & 7937 & 271 & 232 & 19\\
    \enddata
\tablecomments{\tiny Summary of the results from our \WS\ single-color and weighted $W3$ and $W4$ excess identification, using the more accurate determination of the $\Sigma_{E_{CL}}$ outlined in Section~\protect \ref{sec:improved_detection}. $\Sigma_{E_{CL}}$ is the threshold $\Sigma_E$ above which we select an excess at a confidence level higher than $CL$.  $CL=99.5\%$ for $W4$ excesses and 98\% for $W3$ excesses. The number of stars in the parent and science samples for the single-color excess searches are those that pass the selection criteria of \p14 (see also Section~\protect \ref{sec:sampledef}). For the weighted-color excess search we have further required valid detections in all of $W1, W2$, and $W3$ (for $W3$ excesses) or in all four \WS\ bands (for $W4$ excesses).
The final debris disk candidates are the subset of excesses that survive visual inspection for contamination. The last column indicates the number of new detections.}
\tablenotetext{a}{Excess significance threshold for single-color excesses  ($\Sigma_{E_{CL}}$) or weighted-color excesses ($\Sigma_{\overline{E_{\rm CL}}}$).}
\end{deluxetable}

    We used the improved confidence threshold determination procedure to search for additional single-color excesses in the same sets of stars and colors ($W1-W4$, $W2-W4$, $W3-W4$, $W1-W3$ and $W2-W3$) as in \p14. We found 29 additional single-color excess candidates. We rejected HIP~104969, and HIP~111136 after visual and automated inspections (Section~\ref{sec:auto_rejunwise}) for line-of-sight contamination, and we rejected HIP~910 on suspicion of it being a spurious detection (see Section~\ref{sec:newdisk_archival}). We are thus left with 26 single-color excess candidates, 18 of which do not have IR excess detections reported in the literature. Of these 18, 17 are newly detected single-color excesses at $W4$ (99.5\% confidence), and one has a significant (98\% confidence) single-color excess only at $W3$, with a marginal excess at $W4$. The excess detection statistics are summarized in Table~\ref{tab:fdr_thresholds}. The newly detected excesses and their $\Sigma_E$ significances are listed individually in Table~\ref{tab:excessstats}. The 3 rejected single-color excess candidates are included in a list of rejected candidates in Table~\ref{tab:rejects}.
    
\clearpage    
\begin{deluxetable}{lcccccccccc}
\tabletypesize{\tiny}
\tablewidth{0pt}
\tablecaption{IR Excess Information for 75 pc Debris Disk Candidates not Identified in \p14 \label{tab:excessstats}}
\tablehead{\colhead{} &\colhead{}& \colhead{} &\colhead{} & \multicolumn{5}{c}{$\Sigma_E$} & \multicolumn{2}{c}{\ES}\\
\cline{5-9} \cline{10-11}\\
\colhead{HIP} & \colhead{Single Color}& \colhead{Weighted} & \colhead{New?} & \colhead{$W1-W4$} & \colhead{$W2-W4$} & \colhead{$W3-W4$} & \colhead{$W1-W3$} & \colhead{$W2-W3$} & \colhead{Weighted} & \colhead{Weighted} \\ 
\colhead{ID} & \colhead{Excess Flag} & \colhead{Excess Flag} & \colhead{($22|12$\micron)} &  \colhead{} & \colhead{} & \colhead{} & \colhead{} & \colhead{} & \colhead{$W4$} & \colhead{$W3$} } 
\startdata
1893   & NNYNN & NN  & Y- & 2.44 & 3.04 & 2.90 & -0.87 & 0.35 & 2.97 & -0.16 \\
2852   & NNYNN & YN  & Y- & 0.72 & 2.28 & 3.07 & -0.97 & -0.21 & 3.05 & -0.60 \\
12198  & NYYNN & YN  & N- & 2.87 & 3.24 & 3.06 & -0.41 & 0.28 & 3.18 & 0.06 \\
13932  & NYNNN & NN  & Y- & 3.05 & 3.14 & 2.52 & 1.83 & 2.54 & 2.90 & 2.61 \\
18837  & NYYNN & YN  & Y- & 2.67 & 3.16 & 3.03 & -0.24 & 0.15 & 3.15 & 0.04 \\
20094  & NYYNN & YN  & Y- & 2.86 & 3.13 & 3.03 & -0.07 & 0.15 & 3.14 & 0.10 \\
20507  & NNNNN & YN  & Y- & 1.63 & 2.21 & 2.85 & 0.57 & 0.46 & 3.08 & 0.66 \\
21091  & NNYNN & YN  & N- & 2.87 & 3.04 & 3.07 & -0.69 & -0.38 & 3.08 & -0.58 \\
21783  & NYUUU & UU  & Y- & 2.86 & 3.21 & \nodata & \nodata & \nodata & \nodata & \nodata \\
21918  & NYNNN & NN  & Y- & 1.05 & 3.11 & 2.42 & -0.86 & 1.07 & 2.72 & 0.58 \\
26395  & YYYNN & YY  & NN & 13.08 & 21.07 & 20.61 & 1.00 & 3.31 & 23.18 & 3.28 \\
39947  & NNYNN & YN  & Y- & 0.83 & 2.55 & 3.07 & -0.55 & 0.29 & 3.20 & 0.04 \\
42333  & NYNNN & YN  & N- & 0.96 & 3.12 & 2.89 & -0.40 & 1.02 & 3.15 & 0.77 \\
42438  & UNYUN & UU  & N- & \nodata & 2.02 & 3.07 & \nodata & 0.71 & \nodata & \nodata \\
43273  & NYNNN & NN  & Y- & 2.69 & 3.09 & 2.63 & 0.08 & 1.28 & 2.82 & 0.96 \\
58083  & NYYNN & YN  & Y- & 3.08 & 3.23 & 3.05 & -0.05 & 0.46 & 3.17 & 0.32 \\
66322  & NNYNN & YN  & Y- & 1.95 & 2.72 & 3.10 & -0.12 & -0.19 & 3.19 & -0.21 \\
67837  & UUYUU & UU  & Y- & \nodata & \nodata & 2.99 & \nodata & \nodata & \nodata & \nodata \\
70022  & NNYNN & NN  & Y- & 1.75 & 2.47 & 2.94 & -0.02 & -0.30 & 3.01 & -0.27 \\
72066  & UUYUU & UU  & Y- & \nodata & \nodata & 2.92 & \nodata & \nodata & \nodata & \nodata \\
73772  & NYYNN & YN  & Y- & 3.03 & 3.14 & 2.99 & 0.17 & 0.18 & 3.14 & 0.21 \\
78466  & NYYNN & YN  & N- & 2.94 & 3.15 & 2.92 & 0.71 & 0.40 & 3.15 & 0.59 \\
85354  & NYNNN & NN  & Y- & 3.10 & 3.19 & 2.73 & 1.01 & 1.74 & 3.00 & 1.70 \\
92270  & NNYNN & NN  & N- & 1.37 & 1.07 & 2.91 & -0.02 & -1.02 & 2.84 & -0.86 \\
100469 & NNYNN & NN  & NN & 1.79 & 1.41 & 2.99 & 0.10 & -1.60 & 2.88 & -1.38 \\
110365 & NYYNN & YN  & Y- & 3.08 & 3.17 & 3.01 & 0.04 & 0.41 & 3.12 & 0.29 \\
115527 & NNYNN & YN  & N- & 1.88 & 2.86 & 3.13 & -0.24 & -0.10 & 3.20 & -0.18 \\
117972 & NNNYN & NN  & -Y & 2.64 & 1.78 & 0.50 & 2.73 & 2.21 & 1.20 & 2.87 \\
\enddata
\tablecomments{The second column indicates the combination of detections from individual colors. Each flag is a five character string that identifies whether the star has a statistically probable (Y) or insignificant (N) single-color excess in the following order: $W1-W4$, $W2-W4$, $W3-W4$, $W1-W3$ and $W2-W3$. Any star can have an unlisted (U) value, indicating that the star was rejected by the selection criteria for that particular color (Section 2.2 in \p14). ``U'' entries correspond to null entries in the corresponding $Wi-Wj$ $\Sigma_E$ column. Column 3 shows a two-character flag to indicate whether the star has a significant weighted-color excess in the following order: weighted $W4$ excess and weighted $W3$ excess. Column 4 lists whether or not the star has a new excess detection in the $W4$ or $W3$ bands (22 or 12\micron), or not.  Dashed entries (``-'') indicate no detected excess in that band.  The last seven columns list the significance of the excess for each color or weighted metric.}
\end{deluxetable}

    \subsection{Defining a New Weighted-Color Excess Metric}\label{sec:metric}
        
        In \p14 and Section~\ref{sec:improved_detection} we identified debris disk-host candidates by selecting stars with individual anomalously red \WS\ $Wi-Wj$ colors, where $i=1,2,3$, $j=3,4$, and $i<j$. However, it may be possible to attain more reliable excess detections at $Wj$ by combining all relevant $Wi-Wj$ colors. Herein we define this new ``weighted-color excess'' metric.
        
        As in Equation~\ref{eq:old_sig}, we first remove the contribution from the photospheric emission. Thus the single-color excess is: 

\clearpage
\begin{deluxetable}{llc}
\tablewidth{0pt}
\tabletypesize{\footnotesize}
\tablecaption{Rejected \WS\ Excesses\label{tab:rejects}}

\tablehead{\colhead{HIP} & \colhead{WISE ID}  & \colhead{Rejection} \\
           \colhead{ID}       & \colhead{}         & \colhead{Reason}}
\startdata
\hline
\multicolumn{3}{c}{\textbf{New Single-Color and Weighted-Color Excesses}} \\
\hline
HIP910   & J001115.82-152807.2 & 2\\ 
HIP13631 & J025532.50+184624.2 & 1 \\ 
HIP27114 & J054500.36-023534.3 & 1 \\
HIP60689 & J122617.82-512146.6 & 1,3 \\
HIP79741 & J161628.20-364453.2 & 1 \\
HIP79969 & J161922.47-254538.9 & 1,3 \\
HIP81181 & J163453.29-253445.3 & 1 \\
HIP82384 & J165003.66-152534.0 & 1 \\
HIP83221 & J170028.63+150935.1 & 1,3 \\
HIP83251 & J170055.98-314640.2 & 1 \\
HIP99542 & J201205.89+461804.8 & 1,3 \\
HIP104969 & J211542.61+682107.2 & 1,3\\
HIP111136 & J223049.77+404319.8 & 1\\
\hline
\multicolumn{3}{c}{\textbf{Previously Identified Single-Color Excesses from \p14}\tablenotemark{a}} \\
\hline
HIP19796\tablenotemark{b} & J041434.42+104205.1 & 3\\ 
HIP20998 & J043011.60-675234.8 & 3\\ 
HIP28498 & J060055.38-545704.7 & 3\\ 
HIP35198 & J071625.22+350102.8 & 4\\ 
HIP60074\tablenotemark{b} & J121906.38+163252.4 & 4\\
HIP63973 & J130634.58-494111.0 & 3,4\\ 
HIP68593\tablenotemark{b} & J140231.57+313939.3 & 3\\
HIP78010 & J155546.22-150933.9 & 4\\ 
HIP79881 & J161817.88-283651.5 & 3\\ 
HIP95793\tablenotemark{b} & J192900.97+015701.3 & 3\\ 
\enddata
\tablecomments{Rejection reasons: \\
    1. Contamination by nearby infrared source based on visual ``by-eye'' inspection.\\
    2. Spurious excess. See Section~\protect \ref{sec:newdisk_archival}.\\
    3. Contaminated by extraneous extended emission based on a significant difference between the $W4$ photocenters in narrow and wide $W4$ apretures (Section~\protect \ref{sec:centroid_calc}).\\
    4. Contaminated by an extraneous point-source based on a significant difference between the $W3$ and $W4$ photocenters (Section~\protect \ref{sec:centroid_calc}).}
\tablenotetext{a}{These rejected excesses were also recovered using our improved single-color detection techniques.}
\tablenotetext{\textbf{b}}{These rejected excesses have been confirmed as debris disk hosts by higher angular resolution \spitzer\ observations. See Section~\protect \ref{sec:reject_fidelity}.}
\end{deluxetable}

\begin{equation}\label{eq:excess_1}
        E[Wi-Wj] = Wi-Wj-W_{ij}(B_T-V_T). 
\end{equation}

\noindent Since we want to use the strength of all possible \WS\ color combinations for band $Wj$, we constructed the weighted average of the color excesses as

\begin{equation}\label{eq:wtavgExcess}
        \overline{E[Wj]} = \frac{1}{A} \sum\limits_{i=1}^{j-1} \frac{E[Wi-Wj]}{{\sigma_{Wi}}^2},
\end{equation}

\noindent where $\sigma_{Wi}$ is the photometric uncertainity of $Wi$ and $j = 3,4$.  Here, $A=\sum\limits_{i=1}^{j-1} \frac{1}{\sigma_{i}^2}$ is a normalization constant. Our definition for the significance $\left(\Sigma_{\overline{E[Wj]}}\right)$ of the weighted-color excess at $Wj$ is the ratio of the weighted average of all color excesses (Equation~\ref{eq:wtavgExcess}) to the uncertainty in the weighted average ($\sigma_{\overline{E[Wj]}}$):

\begin{eqnarray}\label{eq:combined_significance}
    \Sigma_{\overline{E[Wj]}} &=& \overline{E[Wj]}/\sigma_{\overline{E[Wj]}}\\
            &=& \frac{\frac{1}{A}\sum\limits_{i=1}^{j-1}\frac{E[Wi-Wj]}{\sigma_i^2}}{\sqrt{\sigma_j^2 + 1/A}} . 
\end{eqnarray}

\noindent The full derivation of this metric can be found in Appendix~\ref{sec:appendix}. We use \ES\ throughout the rest of the paper as shorthand for the significance of the weighted-color excess for either $W3$ or $W4$, as appropriate, and $\Sigma_{E}$ as shorthand for the significance of the single-color excess when the discussion does not refer to any specific color.

    \subsection{Weighted-Color Excesses}\label{sec:weighted_excesses_results}

    We extend the same procedure used to identify stars with single-color excesses in Section~\ref{sec:improved_detection} to search for optimally weighted-color excesses in $W3$ or $W4$ using Equation~\ref{eq:combined_significance}. When discussing weighted excesses, we denote the confidence threshold as $\Sigma_{\overline{E_{CL}}}$. We plot the \ES\ distributions as solid red histograms for both $W3$ and $W4$ in Figure~\ref{fig:ColorDist}. The positive wings of the uncertainty distributions, defined analogously to those for the single-color uncertainty distributions, are shown as dashed blue histograms. The $\Sigma_{\overline{E_{CL}}}$ threshold is shown as the vertical dotted green line. We claim that a star has a significant weighted-color excess if its \ES\ $\geq \Sigma_{\overline{E_{CL}}}$.
    
    We identify 6 stars with 98\% significant weighted $W3$ excesses within 75~pc of the Sun, among which we expect $2\% \times 6 = 0.12$ to be false positives. We identify 187 stars with 99.5\% significant weighted $W4$ excesses within 75~pc of the Sun, among which we expect $0.5\% \times 187 = 0.94$ to be false positives. These FDRs only take into account the probability of detecting an excess due to random noise, and do not filter out real excesses that may be caused by other astrophysical contaminants (e.g., IR cirrus or unresolved projected companions).
    
    As with the single-color excess candidates (Section~\ref{sec:improved_detection}), we performed visual and automated inspection of the \WS\ images to determine contamination.  None of the six weighted $W3$ excesses were deemed to be contaminated, while 14 of the 187 weighted $W4$ excess sources were found to be contaminated.  Three of these stars, HIP~69281, HIP~69682, and HIP~106914 were rejected in \citet{Patel2015} due to contamination by nearby background sources. Ten of the 14 have single-color excess detections that were already rejected as debris disk candidates in either \p14 or \citet{Patel2015} and again in Section~\ref{sec:improved_detection}. The remaining one, HIP~111136, is a new weighted $W4$ excess candidate, and was also detected by our improved single-color detections in Section~\ref{sec:improved_detection}, but had not been identified as a single-color excess in \p14. However, we rejected it as its $W4$ images reveal line-of-sight IR cirrus contamination.
    
    \begin{figure}
    \centering
    \includegraphics[scale=0.5]{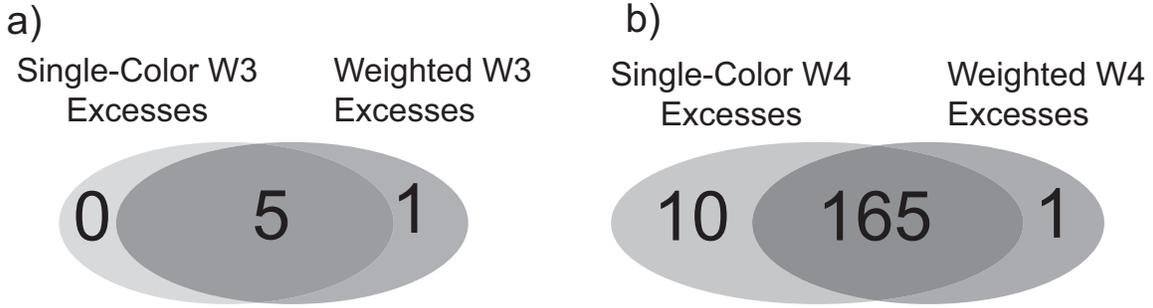}
    \caption{Venn diagrams comparing the candidate excesses from the single-color excess selection (left circles; Section~\ref{sec:improved_detection}) and the weighted-color excess selection (right circles; Section~\ref{sec:weighted_excesses_results}). For the $W3$ comparison in panel (a) the single-color excess set includes only stars with good quality photometry in all of $W1$, $W2$ and $W3$ bands.  For the $W4$ comparison in panel (b) the single-color excess set includes stars with good quality photometry in all four \WS\ bands.\label{fig:Venn}}
    \end{figure}

    Except for HIP~69281, HIP~69682, and HIP~106914, we list the remaining 11 rejected sources in Table~\ref{tab:rejects}. In section~\ref{sec:unwise_rejectmethod}, we remove an additional seven stars, leaving us with 166 weighted $W4$ excess stars (Table~\ref{tab:fdr_thresholds}). Figure~\ref{fig:Venn} shows the relation and overlap between the single-color and weighted-color $W3$ and $W4$ excess detections.

\section{Automated Rejection of Contaminated Stars Using Reprocessed WISE Images}\label{sec:auto_rejunwise}

    \WS\ offers higher angular resolution than \iras.  However, source photometry is still prone to contamination by unrelated astrophysical sources seen in projection.  Possible contaminants may include nearby point sources at angular separations comparable to the sizes of the \WS\ $W3$ and $W4$ point-spread functions (PSFs).  Even if the All-Sky Catalogue provides resolved photometry for such objects, the deblending algorithm may introduce systematic errors in the flux that are not characteristic of isolated point sources.  Other possible contamination can be caused by nearby extended emission: e.g., from interstellar cirrus or from the PSF wings of a nearby bright source.  We expect that both types of contamination may manifest themselves in discrepant source positions: either between the $W3$ and $W4$ images, or among $W4$ positional measurements that use different photocentering region sizes.

    Neither the \WS\ All-Sky Survey Catalog nor the AllWISE Catalog list astrometric positions in each of the separate bands.  Therefore, we downloaded the co-added $W3$ and $W4$-band images for all stars in our parent sample to measure their band-specific positions.  As we describe below, we used images with the native \WS\ angular resolution rather than the smoothed, $\sqrt{2}\times$ broader images accessible from the \WS\ All-Sky Survey or AllWISE data releases.

    \subsection{Using {\sc unWISE} Images to Identify Contaminants}\label{sec:centroid_calc}

    Instead of using the co-added and mosaicked `Atlas' images from the \WS\ All-Sky Survey, we used the higher angular resolution {\sc unWISE} images, which can be retrieved from the {\sc unWISE} image service\footnote{\url{http://unwise.me}} \citep{Lang2014}.  In the official All-Sky Survey and AllWISE data releases, the final images were created by stacking individual exposures and then convolving each stack with a model of the detector's PSF.  In contrast, the {\sc unWISE} images were created by eliminating the  final convolution step, thus preserving the original \WS\ resolution  \citep{Lang2014}. Hence, the {\sc unWISE} PSF is a factor of $\sqrt{2}$ narrower than for the All-Sky Catalog images ($\sim$6.0$\arcsec$ vs.\ $\sim$8.5\arcsec\ at $W1$, $W2$, $W3$ and $\sim$12\arcsec\ vs.\ $\sim$17.0\arcsec\ at $W4$). 
    
    We downloaded 150\arcsec\ $\times$ 150\arcsec\ postage-stamp $W3$ and $W4$ images from the {\sc unWISE} website for all of our excess candidates, each centered on the stellar coordinates at the mean \WS\ observational epoch. We also downloaded images for the 16960 \p14 parent sample stars: \hip\ main sequence stars within 120~pc. This sample is the union of all the stars that comprised the parent samples for the five different color excess searches in \p14: $W1-W3$, $W2-W3$, $W1-W4$, $W2-W4$, and $W3-W4$. We use this amalgamated parent sample as a basis for determining which candidate excess stars have statistically significant positional discrepancies.
    
    We explored two independent ways to automatically detect unrelated contamination: one primarily for point sources and one for extended sources. We hypothesized that unrelated point-source contaminants can be identified through significant positional offsets between the centroids of the $W3$ and $W4$ {\sc unWISE} images. These would represent cases where the catalogued $W4$ excess is caused by the contaminating source, which would then likely have a much redder $W3-W4$ color than the target star.  The $W4$ centroid of the target star would then be shifted away from the $W3$ centroid, in the direction of the contaminating object. We extracted $W3$ and $W4$ centroid positions for the parent sample stars from the {\sc unWISE} postage stamps.  We denote these as $\vec{r}_{_{W3}}$ and $\vec{r}_{_{W4}}$, respectively. The centroid positions were obtained from 2D Gaussian fits to the pixel values in a 3.06~pixel (8.42\arcsec) radius aperture, with a Gaussian of $\sigma=1.02$~pixels. The $\sigma$ value was chosen to yield a full width at half maximum (FWHM) of 2.40~pixels (6.60\arcsec), slightly larger than the FWHM of the $W3$ {\sc unWISE} PSF.
    
    We also hypothesized that extended-source contaminants could be identified by comparing the $W4$ centroid calculated in an $r=3.06$~pixel (8.42\arcsec) aperture to a $W4$ centroid calculated in a wider $r=10.0$ pixel (27.5\arcsec) aperture (extending out to the second Airy minimum).  These would correspond to cases where a star is projected on a background of interstellar cirrus.  The smaller-aperture centroid would be dominated by the stellar PSF, while the wider-aperture centroid would be weighted more strongly by the spatial distribution of the cirrus.  If the cirrus surface brightness distribution is uneven, that would generally result in a systematic offset between the narrow- and wide-aperture centroids.  As before, we extracted $W4$ centroid positions for the parent sample stars from the {\sc unWISE} postage stamps.  We denote the $W4$ wide-aperture centroids as $\vec{r}_{_{W4},_{wide}}$. 

    Altogether, we aim to automatically identify contaminants based on large offsets between the $W3$ and $W4$ image centroids ($\vec{r}_{_{W3,W4}} = \vec{r}_{_{W3}} - \vec{r}_{_{W4}}$), or between the $W4$ image centroids calculated from narrow vs.\ wide apertures ($\vec{r}_{_{W4,W4}} =\vec{r}_{_{W4}} - \vec{r}_{_{W4},_{wide}}$).  We can set the threshold for contamination in our science sample by studying the distribution of positional offsets for the parent sample.  We can then mark as contaminated all science sample stars with offsets larger than the chosen threshold for either of the methods.
    
     \subsection{Rejecting Astrometric Contaminants}\label{sec:unwise_rejectmethod}

    \begin{figure}
    \centering
    \includegraphics[angle=90,scale=0.85]{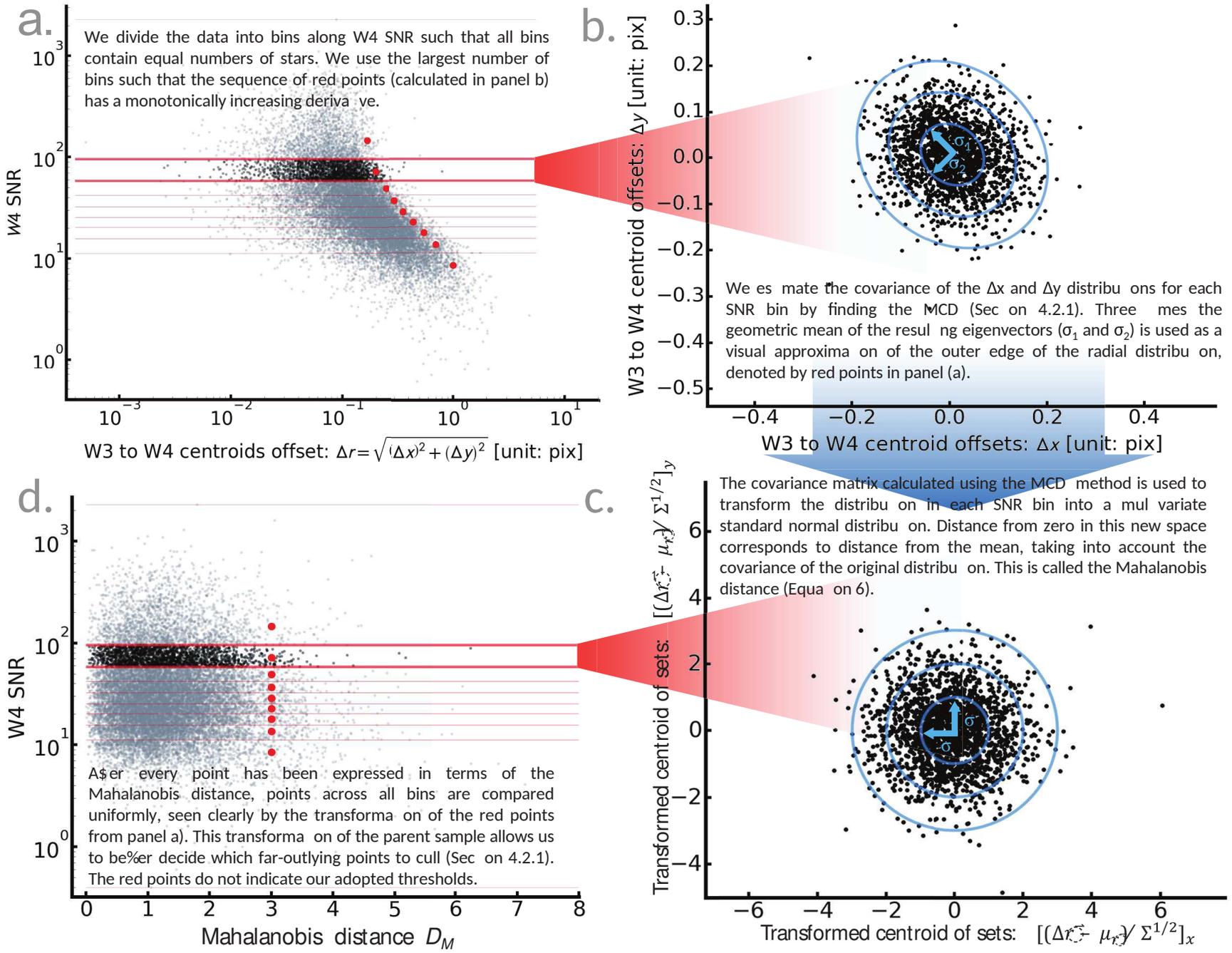}
    \caption{An illustration depicting the steps taken to derive the Mahalanobis distances of the astrometric offsets for each star in the parent sample, as described in \S~\ref{sec:eliminating_dependencies}.\label{fig:mnd_illustration}}
    \end{figure}

    The automated contamination checking approach outlined in the preceding Section~\ref{sec:centroid_calc} needs to take into account two considerations.  First, the positional uncertainty of an object depends on its signal-to-noise ratio (SNR).  Consequently, the distribution of the $\vec{r}_{_{W3,W4}}$ and $\vec{r}_{_{W4,W4}}$ centroid offsets varies as a function of SNR. Therefore, the rejection threshold needs to depend on SNR.  Second, the  positional $x$ and $y$ uncertainties are correlated in pixel coordinates because the \WS\ PSF is not circularly symmetric.  For example, the $W3$ PSF has (post-convolution) major and minor axes of 7.4\arcsec and 6.1\arcsec\footnote{See Table 1 in Section IV.4.c.iii.1 of the All-Sky Explanatory Supplement; \url{http://wise2.ipac.caltech.edu/docs/release/allsky/expsup/sec4\_4c.html\#psf}}.  Consequently, the distribution of the centroid offsets $\vec{r}_{_{W3,W4}}$ and $\vec{r}_{_{W4,W4}}$ will not be centrally symmetric, and their $\Delta x$ and $\Delta y$ projections onto pixel coordinates will be correlated.  Generally, the $\Delta x$ and $\Delta y$ distributions will follow different degrees of correlation as a function of SNR.
    
    We illustrate these two considerations for the $\vec{r}_{_{W3,W4}}$ centroid offsets in panels (a) and (b) of Figure~\ref{fig:mnd_illustration}. The bean-like cloud of data points in Figure~\ref{fig:mnd_illustration}a shows a clear trend for a widening distribution of $\Delta r^2 = \Delta x^2 + \Delta y^2$ variances in the $\vec{r}_{_{W3,W4}}$ centroid offsets at lower $W4$ SNRs.  The elongated 2D distribution of $\Delta x$ vs.\ $\Delta y$ in Figure~\ref{fig:mnd_illustration}b shows the covariance expected from the centrally asymmetric shape of the \WS\ PSF.

    \subsubsection{Eliminating SNR and Covariance Dependencies in the Astrometry}\label{sec:eliminating_dependencies}

    The covariance of the $\Delta x$ and $\Delta y$ offsets at any $W4$~SNR means that we cannot determine the significance of a star's astrometric offset by simply calculating $\Delta r^2 = \Delta x^2 + \Delta y^2$. Instead, we require a distance statistic that is independent of the covariance among $\Delta x$ and $\Delta y$.  In addition, because the covariance of the $\Delta x$ and $\Delta y$ offsets depends on SNR, the covariance matrix must be calculated at different $W4$~SNRs. 
    
    We start by binning our parent sample in $W4$~SNR bins in the $W4$~SNR vs.\ $|\vec{r}| = \Delta r$ space. The binning is illustrated in Figure~\ref{fig:mnd_illustration}a. The bins are not equally spaced, but are instead chosen such that all bins contain an equal number of stars, which in turn ensures that there are no under-represented bins. To determine the optimal number of bins, we first start with a small number (e.g., 4) of bins, and in each bin calculate the geometric mean of the variances along the principal axes of the 2D $\Delta x$ vs.\ $\Delta y$ distribution: i.e., the eigenvalues of the covariance matrix. The geometric mean approximates what the (joint) variance would be if the positional offsets in $\Delta x$ and $\Delta y$ were uncorrelated and had equal variance. The geometric means of the $\Delta x^2$ and $\Delta y^2$ variances for each bin are shown as red points in Figure~\ref{fig:mnd_illustration}a, where they are multiplied by 3 for illustrative purposes. We then increased the number of bins until the geometric means for all bins stopped forming a sequence that had a monotonically increasing derivative. For our analysis, we thus used nine equally populated bins. We expect the relationship between SNR and astrometric offsets to be smooth, and using more than nine bins results in a jagged approximation.
    
    We then need to determine how the empirical distribution of the geometric means of the $\Delta x^2$ and $\Delta y^2$ variances can be used to set a probability threshold for contamination. Each population of $|\vec{r}|$ offsets in the $W4$ SNR bins is comprised of an underlying statistically random population and an outlier population. The covariance matrix of the $\Delta x$ and $\Delta y$ offsets must be calculated for the statistically random sample while being insensitive to the presence of outliers. To this end, we adopt the minimum covariance determinant \citep[MCD;][]{Rousseeuw1999} method. 
    
    The MCD method is optimized to selectively ignore data that are significantly distant from the center of the distribution, such that the determinant of the resulting covariance matrix $\Sigma_{\Delta x, \Delta y}$ is minimized.
    Figure~\ref{fig:mnd_illustration}b illustrates the covariance ellipses calculated by the MCD technique, for a given $W4$ SNR bin.
    
    Finally, we adopt a dimensionless distance metric, the Mahalanobis distance \citep{Mahalanobis1936}, to represent all astrometric offset measurements. Doing so allows us to normalize over the differences in the lengths of the eigenvectors of the $\Sigma_{\Delta x, \Delta y}$ covariance matrices among the $W4$ SNR bins. We calculate the Mahalanobis distance $D_{M}$ using a matrix multiplication of the observed offset $\Delta r=(\Delta x,\Delta y)$ and the distribution's covariance matrix ($\Sigma_{\Delta x, \Delta y}$):
    
    \begin{equation}\label{eq:mnd}
    D_M^2=\mathbf{r}^{T}\Sigma^{-1}_{\Delta x, \Delta y}\mathbf{r}.
    \end{equation}

    The calculation of the Mahalanobis distance is the multi-dimensional equivalent of subtracting the mean of the distribution and dividing by the standard deviation.  In essence, we are  performing two separate transformations to the 2-D $\Delta x$ and $\Delta y$ offset distributions: a rotation and scaling. The rotation is dictated by the eigenvectors of the covariance matrix $\Sigma^{-1}_{\Delta x, \Delta y}$, while its eigenvalues determine the magnitude of the scaling. The transformed 2-D offset distribution is then centrally symmetric, with the Mahalanobis distance $D_M$ describing the radial distance of each data point from the origin in units of the standard deviation of the distribution (see Figure~\ref{fig:mnd_illustration}c--d).

    We calculate the Mahalanobis distances separately for each bin, since the covariance matrices differ. Figure~\ref{fig:mnd_illustration}c shows how the 2-D $\Delta x$ vs.\ $\Delta y$ distribution for a given $W4$ SNR bin is transformed after being decorrelated and normalized (by dividing out the square root of the covariance matrix). Figure~\ref{fig:mnd_illustration}d shows the final version of the $W4$ SNR vs.\ $|\vec{r}|$ distribution, where the $|\vec{r}|$ offsets have been expressed in terms of the dimensionless Mahalanobis distances. The Mahalanobis distance distributions are identical (by design) across all bins, which allows us to set a uniform threshold for rejecting positional outliers.

        \subsubsection{Adopting A Uniform Rejection Threshold}\label{sec:rej_threshold}
    
    In the absence of contamination by nearby sources, the centroids of the majority of the stars would be distributed according to a multivariate normal distribution.  Consequently, the Mahalanobis distances would follow a $\chi^2$ distribution of two degrees of freedom. We aim to separate the population of uncontaminated stars from the outlier population of contaminated stars whose centroids are offset because of nearby emission.  As an estimate of the uncontaminated population, we select all stars with $D_{M}<2$. Since the population of uncontaminated stars dominates at such small offsets, and since the spatial distribution of its centroid offsets is expected to be narrower, we expect the set of $D_{M}<2$ stars to not be significantly affected by contamination.  We denote $f(x)$ to be the probability density function of the $\chi^2$ distribution with two degrees of freedom representing the uncontaminated population, while $N_{D_M<2}$ is the number of stars in this population. Thus, the uncontaminated distribution can be represented using the empirical data and scaled such that

    \begin{equation}\label{eq:scale_mnd}
    A\int_{0}^{2}f(x)dx= N_{D_M<2} ,
    \end{equation}

    \noindent where $A$ is the normalization factor. 
    
    We then calculate $A$ from Equation~\ref{eq:scale_mnd} and use it to compare the empirical $D_M$ distribution for the centroid offsets to the expectation $A f(x)$ for an uncontaminated distribution. We estimate the fraction of stars within a certain $D_M$ that are expected to be uncontaminated by calculating the negative predictive value (NPV) as a function of $D_M$. If we set a threshold $D_{M_0}$ beyond which we reject stars as astrometrically contaminated, then the NPV is defined as: 
    
    \begin{equation}\label{eq:npv}
    NPV=\frac{A\int_{0}^{D_{M_0}}f(x)dx}{N_{D_M<D_{M_0}} }.
    \end{equation}

    In our case, we set the NPV $=99.5\%$ and solve Equation~\ref{eq:npv} for $D_{M_0}$ by calculating the intersection of the right and left hand side of Equation~\ref{eq:npv}. We find $D_{M_0}$ thresholds of 3.63 and 3.28 for the $W3$ vs.\ $W4$ and $W4$-narrow vs.\ $W4$-wide analyses, respectively.  
    
    \begin{figure}
    \centering
    \includegraphics[width=3.2in]{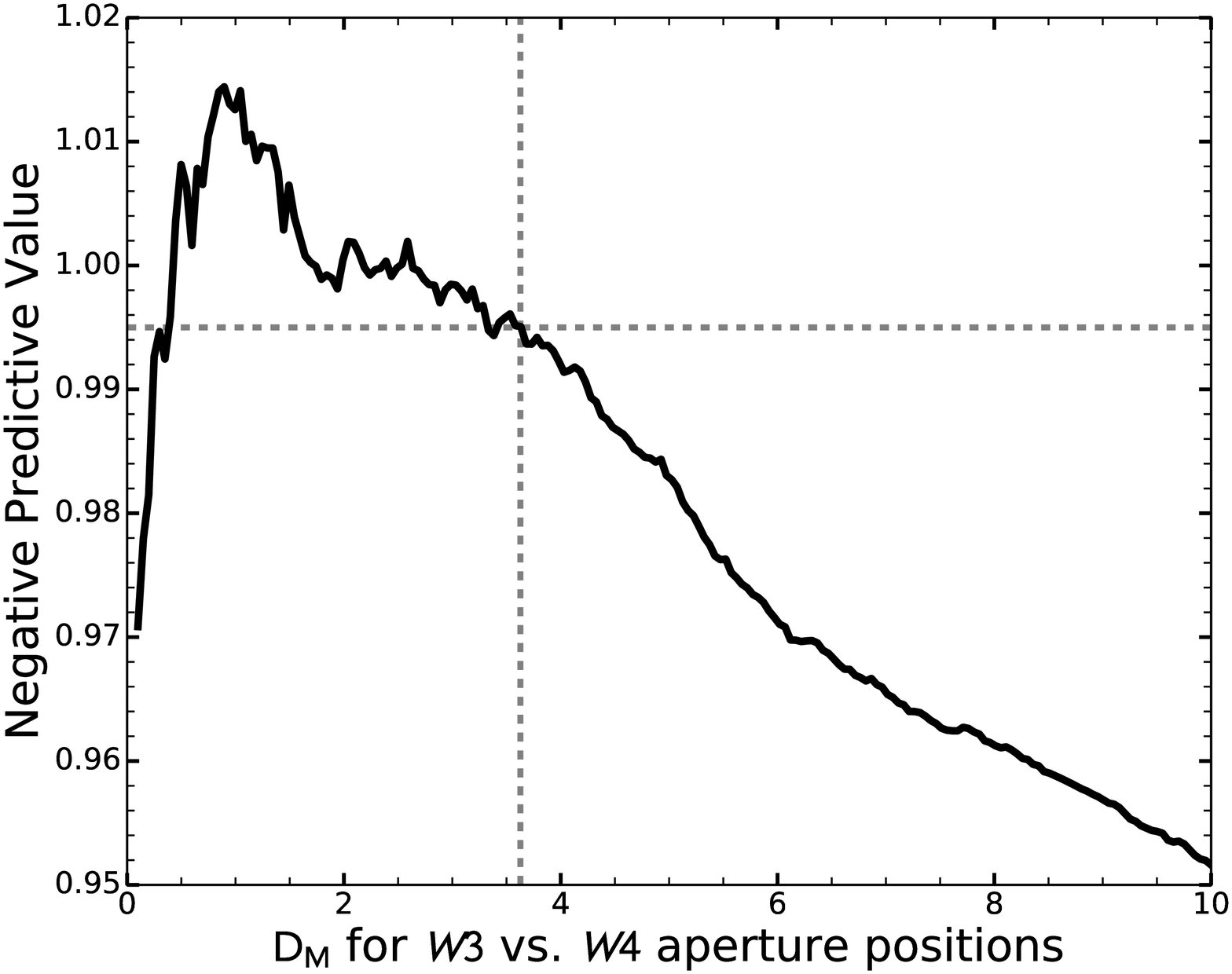}
    \includegraphics[width=3.2in]{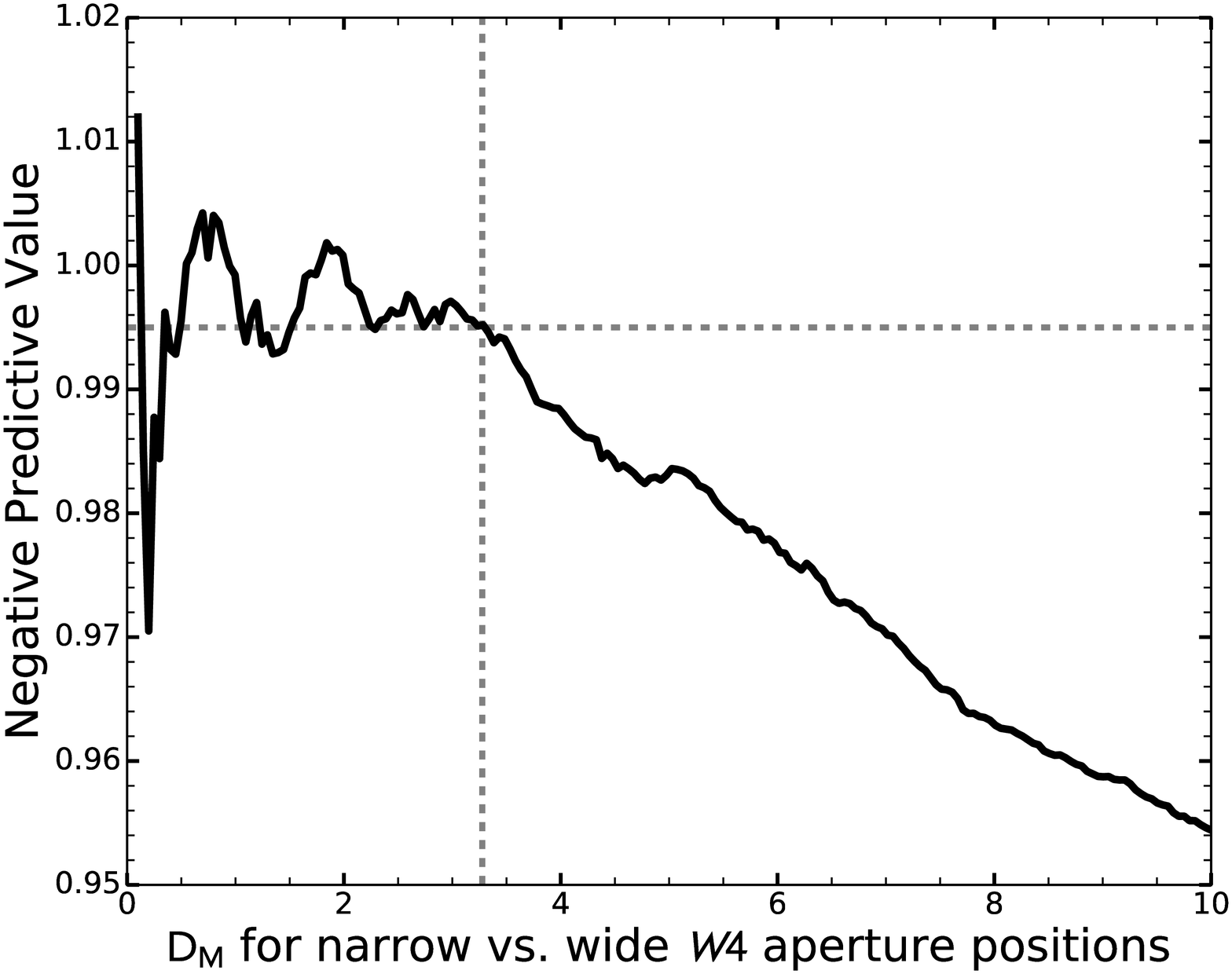}
    \caption{The NPV distributions of the 120~pc parent sample stars as a function of the Mahalanobis distances between their ($x, y$) positions in {\sc unWISE} images.  The horizontal dashed line is set at NPV=99.5\%. The vertical dashed line indicates $D_{M_0}$, solved from equation~\ref{eq:npv}. Stars with $D_M>D_{M_0}$ (3.63 and 3.28 for the $W3$ vs.\ $W4$ and $W4$-narrow vs.\ $W4$-wide analyses, respectively)} and NPV $<0.995$ are rejected as astrometric outliers. \textit{Left:} NPV distribution for $W3$ vs.\ $W4$ offsets.  \textit{Right:} NPV distribution for offsets between the narrow (2.5~pix) radius and wide (10~pix) radius apertures in $W4$.\label{fig:npv}
    \end{figure}
    
    Figure~\ref{fig:npv} shows the NPV distributions for the two analyses, with the NPV $=99.5\%$ $D_{M_0}$ thresholds marked with vertical lines. Should the distribution of centroid offsets at $D_M<2$ have been ideally represented by a $\chi^2$ distribution with two degrees of freedom, the NPV distributions would start at unity at $D_M=0$ and monotonically decrease toward larger values of $D_M$.  However, since we are dealing with a real data set, the NPV distributions are noisy at small $D_M$ (fewer data points) and become monotonic only at larger $D_M$.  Therefore, while there are several possible $D_M$ values at which NPV $=99.5\%$, we retain the largest one as our threshold $D_{M_0}$. We reject candidate excesses with Mahalanobis distances above these thresholds. 
    
\begin{figure}
    \centering
    \includegraphics[scale=0.4]{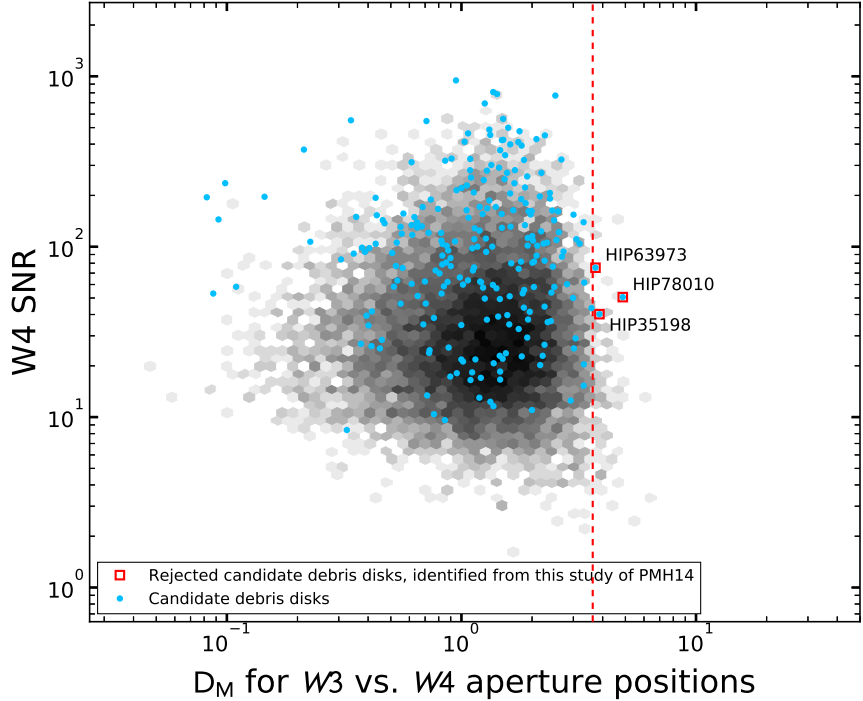}
    \caption{$W4$ SNR vs.\ Mahalanobis distance between the $W3$ and $W4$ {\sc unWISE} centroids (see Sections~\ref{sec:centroid_calc}--\ref{sec:unwise_rejectmethod}). The black/gray density cloud represents the density of 16927 \hip\ 120~pc parent sample stars.  The light-blue dots represent the candidate excess stars. The vertical black-dotted line represents the NPV=99.5\% threshold for rejecting astrometrically contaminated excesses. The {\sc unWISE} images for the rejected stars are shown in Figure~\ref{fig:w3w4_postagestamps}.\label{fig:w3w4_snrvsep}}
    \end{figure}

\begin{figure}
    \centering
    \includegraphics[scale=0.4]{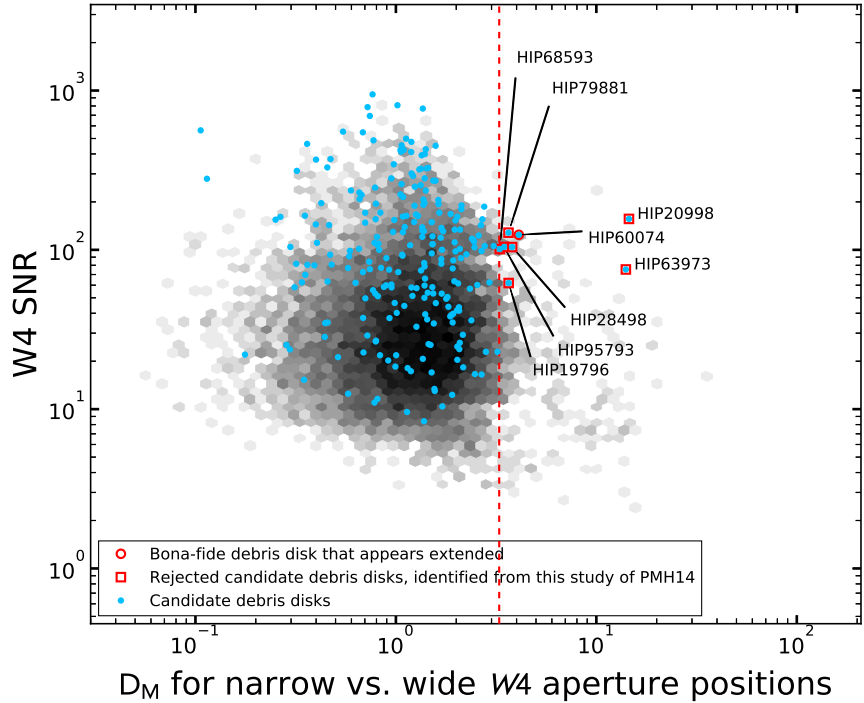}
    \caption{$W4$ SNR vs.\ Mahalanobis distance between the $W4$ {\sc unWISE} centroids in narrow (2.5~pix) and wide (10~pix) apertures (see Sections~\ref{sec:centroid_calc}--\ref{sec:unwise_rejectmethod}). The black/gray density cloud represents the density of 16927 \hip\ 120~-pc parent sample stars.  The light-blue dots represent the candidate excess stars. The vertical black-dotted line represents the NPV=99.5\% threshold for rejecting astrometrically contaminated excesses. The contaminated objects include eight candidate debris disk excesses identified in this study, and are marked with open square symbols. The {\sc unWISE} images for the rejected excesses are shown in Figure~\ref{fig:w4w4_postagestamps}.\label{fig:w4w4_snrvsep}} 
    \end{figure}

    Figures~\ref{fig:w3w4_snrvsep} and \ref{fig:w4w4_snrvsep} show the distribution of the Mahalanobis distances with respect to the $W4$ SNRs for both analyses. We find that three of the candidate excesses, associated with HIP~35198, HIP~63973, and HIP~78010 are rejected because of large $W3$-to-$W4$ centroid offsets (Figure~\ref{fig:w3w4_snrvsep}), and eight candidate excesses are rejected because of large centroid offsets between the narrow and wide $W4$ apertures (Figure~\ref{fig:w4w4_snrvsep}). Only HIP~63973 is rejected by both techniques. All of these rejected stars were previously identified in \p14 as single-color $W4$ excesses and except for HIP~19796, HIP~20998, and HIP~28498 (due to ``bad'' $W1$ and $W2$ photometry), were also identified as weighted-$W4$ excesses in this study. In the following, we address the reliability of our automatic rejection approach. 
    
  \subsection{Rejection Fidelity}\label{sec:reject_fidelity}
  
    We would like to determine whether stars rejected by our automated positional analysis of {\sc unWISE} images are indeed contaminated. The expectation is that if an extraneous point or extended source can randomly offset the centroid positions (and hence contaminate the photometry) of a star, then the fraction of rejected (contaminated) stars among our candidate excesses should be higher than the fraction of rejected stars in an the non-excess portion of the science sample. This is because if a contaminating source is bright enough to influence the photocenter of the star, it is likely to increase the flux of the star as well. 
    
    To this end, we compare the fraction of astrometrically rejected stars in two complementary subsets of the science sample. On one hand we consider the population of 271 candidate excesses before any visual or automated rejection, and on the other hand we take its complement of 7666 non-excess stars.  We use Welch's t-test to determine whether the fractions of stars rejected from each subset by the centroid checks are significantly different from each other. Thus, this test will tell us whether the null hypothesis can be rejected. Specifically, the null hypothesis is that the means of the rejected and complementary science samples are equal.
    
    The result from this test yielded a $p$-value of 0.025, indicating that the probability of observing the difference in the means of the two populations, assuming they are the same, is 2.5\%. With this, we can reject the null hypothesis and claim that the mean of the two populations are not equal. In other words, though this test does not determine whether all stars astrometrically rejected excesses are contaminated, it does tell us that the astrometric rejection technique is indeed preferentially selecting stars that are selected as candidate excesses. 
    
    Our automated checks for contamination by nearby point or extended sources are sensitive to systematic offsets as small as 0.2~pix (0.6\arcsec) at SNR $>100$.  This corresponds to a small fraction of the FWHM of the raw {\sc unWISE} PSF: a tenth at $W3$ or a twentieth at $W4$. The human eye may be challenged at discerning such small offsets.  Nonetheless, it is always instructive to perform a visual inspection of the actual images of the rejected sources. 

\begin{figure}
    \centering
    \includegraphics[width=.7\textwidth]{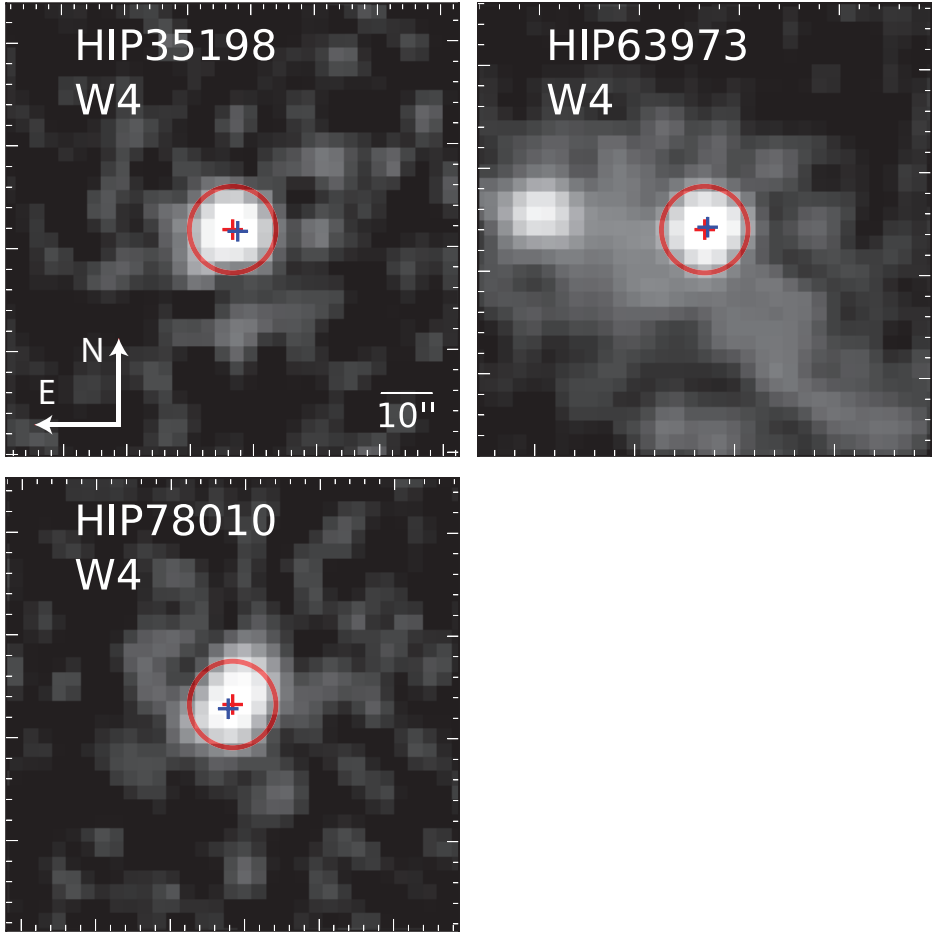}
    \caption{44\arcsec$\times$44\arcsec\ {\sc unWISE} $W4$ postage-stamp images stars rejected by our point source contamination check because of significant offsets between the $W3$ and $W4$ narrow aperture centroids (3.06~pixels or 8.42\arcsec). The red and blue crosses show the centroid locations calculated from the $W4$ and $W3$ images, respectively. They are over-plotted on only the $W4$ images for comparison. The red circles denote the 3.06~pixels (8.42\arcsec) radius aperture used to calculate the centroid position in both bands. \label{fig:w3w4_postagestamps}}
    \end{figure}

\begin{figure}
    \centering
    \includegraphics[width=\textwidth]{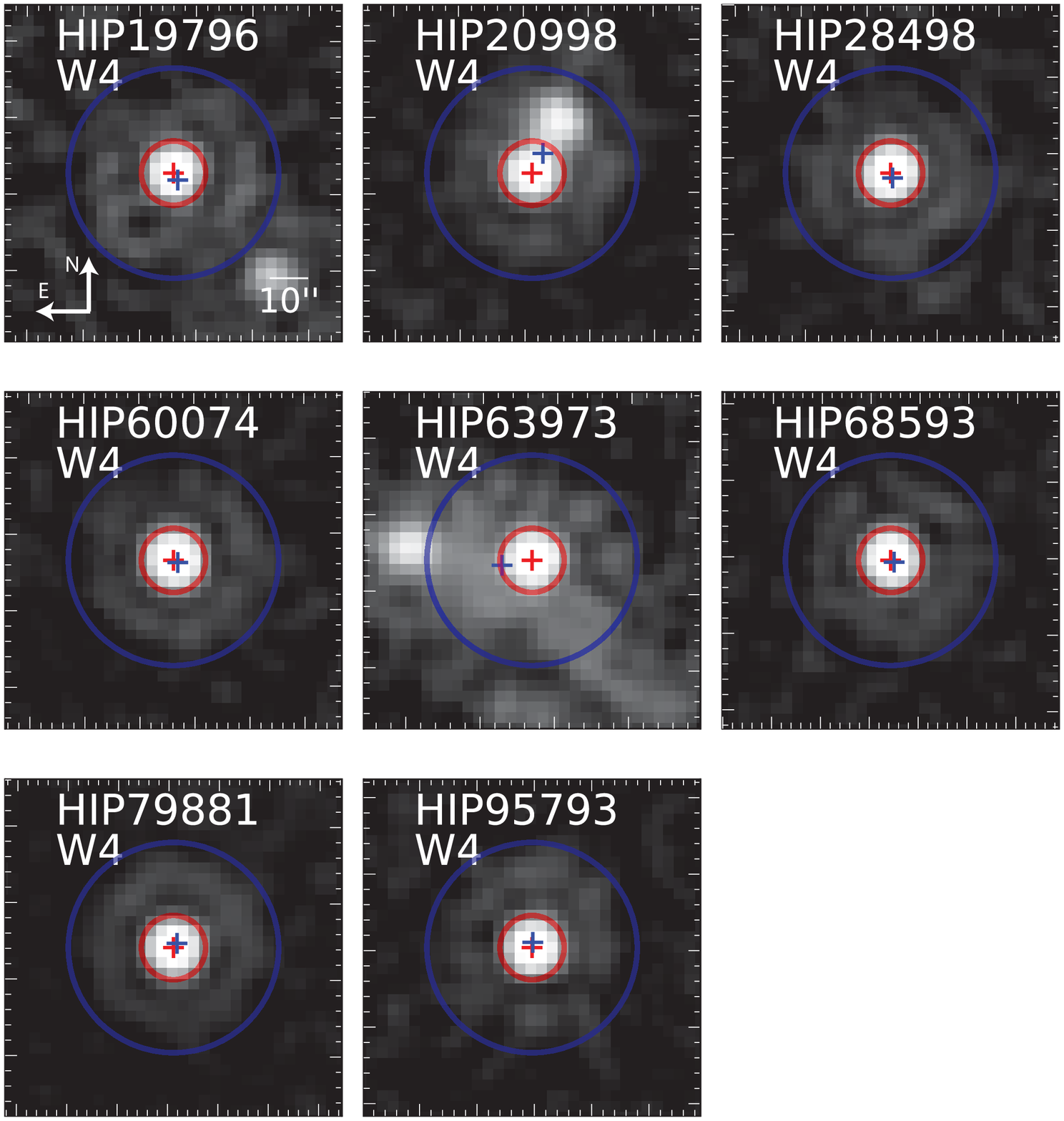}
    \caption{44\arcsec$\times$44\arcsec\ {\sc unWISE} $W4$ postage-stamp images of stars rejected by our extended source contamination check because of significant offsets between the $W4$ centroids in narrow (red circle, 3.06~pixels or 8.42\arcsec) and wide (blue circle, 10~pixels or 27.5\arcsec) apertures. The red and blue crosses in each image are the centroid locations calculated from their respective colored apertures.\label{fig:w4w4_postagestamps}}
    \end{figure}

    Figures \ref{fig:w3w4_postagestamps} and \ref{fig:w4w4_postagestamps} show postage-stamp {\sc unWISE} images of the rejected candidate excesses. Some of the automatically rejected sources clearly show contamination from nearby emission in the {\sc unWISE} images.  This is the case for two of the candidates---HIP~20998, and HIP~63973---rejected by the $W4$ narrow vs.\ wide aperture centroid comparison (Figure~\ref{fig:w4w4_postagestamps}).
    
    
    Conversely, the visual case for rejecting the remaining candidates is less clear cut. For instance, HIP~79881 does not appear to be contaminated by extended cirrus based on its zoomed-in {\sc unWISE} postage stamp image. However, the All-Sky Atlas images show the star to be partially contaminated by cirrus. Indeed, \citet{Rebull2008} discusses the lack of a \textit{Spitzer}/MIPS 24\micron\ excess, attributing previous \iras\ detections with the blending of the source and IR cirrus. In addition, \citet{Riviere-Marichalar2014} do not detect an excess at 70\micron.  These two studies corroborate our rejection of this excess detection. The images of HIP~35198, and HIP~78010, which possess the largest $D_M$ based on their $W3$-to-$W4$ centroids (seen in Fig.~\ref{fig:w3w4_postagestamps}), show some tenuous extended emission at $W4$, as may HIP~28498 and HIP~95793 (Figure~\ref{fig:w4w4_postagestamps}). However, no visible contamination can be seen around most excess candidates rejected at $D_M \lesssim 4$.
    
    Notably, four of the rejected candidate excesses, associated with HIP~19796 \citep{Urban2012},  HIP~68593 \citep{Zuckerman2004, Rhee2007, Carpenter2009, Chen2014}, HIP~95793 \citep{Su2006, Draper2016}, and HIP~60074 \citep{Ardila2004}, have been established as debris disk hosts, and are confirmed in higher angular resolution observations by \spitzer. The latter, HIP~60074 (HD~107146), is a well-known cold debris disk that has been spatially resolved in scattered light by the {\it Hubble Space Telescope} \citep{Ardila2004} and in the submillimeter by the Atacama Large Millimeter Array \citep[ALMA;][]{Ricci2015}.  In our analysis of the narrow- vs.\ wide-aperture $W4$ centroids it sits slightly beyond the $D_{M_0}$ threshold, below which it would be considered uncontaminated. We note that the centroid offset for this star is $\Delta r_{W4} = 1.26$\arcsec\ in the southwest direction. \citet{Ardila2004} identified a faint background spiral galaxy roughly 6\arcsec\ from HIP~60074 in the same direction as this offset. The position of the galaxy places it within the \WS\ $W4$ beam. The offset between the narrow- and wide-aperture $W4$ centroids, and the $W4$ flux of HIP~60074, may thus be affected by 22$\micron$ emission from the background galaxy. No such projected contaminants are known for the other three previously known debris disks that are rejected by our centroid offset analysis.
    
    It is very likely that some of the stars rejected by the centroid offset comparisons, and for which contamination cannot be visually discerned, have bona-fide IR excesses from debris disks.  Nonetheless, we retain the centroid checks as an unbiased and objective indicator of possible IR flux contamination. Our contamination thresholds are established empirically, from the larger parent sample.  If a contaminant is well blended with the stellar PSF, the centroid offset may be the only reliable way to identify it.
    
    We also note that some of the stars that we reject upon visual inspection are not identified as contaminated by the automated centroid offset comparisons. Among the twelve visually rejected stars in Table~\ref{tab:rejects} (rejection reason equal to 1), seven (HIP~13631, HIP~27114, HIP~79741, HIP~81181, HIP~82384, HIP~83251, HIP~111136) were not identified as being contaminated by our astrometric rejection method. Upon comparing the Atlas and {\sc unWISE} images for each of these seven stars, we find visual differences in the structure of the cirrus, as the {\sc unWISE} images show cirrus which is less pronounced. This is caused mainly by the different smoothing kernels used between the Atlas and {\sc unWISE} service. Thus, one of two explanations are plausible. The first is that our rejection technique has not been fully customized to detect extended cirrus emission below a certain threshold, or more likely, that we are being conservative in our assessment of what is contaminated from a subjective visual inspection.

\section{RESULTS}\label{sec:results}

    Our improved \WS\ IR excess identification procedure has uncovered 29 candidate excesses that we did not report in \p14. In Section~\ref{sec:newdisk_archival} we argue that one of these excesses, associated with HIP~910, is likely spurious, which leaves 28 candidate excess identifications not reported in \p14. These are the 28 excesses whose detection specifics are listed in Table~\ref{tab:excessstats}. Nineteen of the 28 excesses are new to the literature, and are addressed in more detail in Section~\ref{sec:new_disks}.
    
    The 28 excesses newly identified by our color-selection methods include  single-color only excesses (12 at $W4$ and one at $W3$),  weighted-color only excesses (one at $W3$ and one at $W4$), and excesses that have both single-color and weighted-color detections (13 at $W4$). An inspection of the single-color excess significances $\Sigma_E$ for each star shows that all of the new detections are fainter (smaller $f_d$ fractional excesses) than those found in \p14: mainly because of the decrease of the $\Sigma_{E_{CL}}$ confidence level in our updated FDR threshold determination (Sec.~\ref{sec:improved_detection}).

    The stellar and dust properties of the 28 candidate excesses are listed in Tables~\ref{tab:stellarparameters} and \ref{tab:diskcandidates}. These parameters are derived from photospheric model fits to the optical and near-IR photometry from the \hip\ catalogue and the Two Micron All-Sky Sky Survey (\mass), using a procedure similar to the one outlined in \p14. The only update with respect to \p14 is that after fitting the optical/IR SED with a photospheric model to determine the best-fit stellar effective temperature, we then scale the model to the weighted mean of the $W1$ and $W2$ fluxes for consistency with our weighted-excess search methodology. However, we note that without additional longer-wavelength observations, our dust temperature estimates are only approximate.
    
    \floattable
    \begin{deluxetable}{llccccccccccccc}
    \tabletypesize{\tiny}
    \tablewidth{0pt}
    \tablecaption{Parameters of Stars with \WS\ Color Excesses Identified Since \p14 \label{tab:stellarparameters}}
    \tablehead{
    \colhead{HIP} & \colhead{WISE} & 
    \colhead{SpT\tablenotemark{a}} & 
    \colhead{Dist.\tablenotemark{b}} &
    \colhead{$T_*$} & \colhead{$R_*$} & 
    \colhead{$\chi^2_*$} & \colhead{$F_{W3}$} & 
    \colhead{$F_{W3,*}$} & \colhead{$F_{W4}$} & 
    \colhead{$F_{W4,*}$} & \colhead{$\Delta_{F_{W3}}/F_{W3}$\tablenotemark{c}}& 
    \colhead{$\Delta_{F_{W4}}/F_{W4}$\tablenotemark{c}}& 
    \colhead{$W1_{corr}$\tablenotemark{d}} & 
    \colhead{$W2_{corr}$\tablenotemark{d}}\\
    \colhead{ID} & \colhead{ID} & \colhead{} &
    \colhead{(pc)} & \colhead{(K)} & 
    \colhead{($R_\sun$)} & \colhead{} &
    \colhead{(mJy)} & \colhead{(mJy)} & \colhead{(mJy)} &
    \colhead{(mJy)} & \colhead{} & \colhead{} & \colhead{(mag)} &
    \colhead{(mag)} 
     } 
    
    \startdata
    1893 & J002356.52-142047.4 & G6V & 53 & 5468 & 1.0 & 1.9 & 48.6$\pm$0.8 & 50.4 & 17.3$\pm$1.1 & 14.0 & -0.036  $\pm$  0.016 & 0.188 $\pm$  0.049 & 6.868$\pm$0.032 & 6.958$\pm$0.023 \\
    2852 & J003606.78-225032.9 & A5m... & 49 & 7448 & 1.6 & 1.4 & 194.2$\pm$2.7 & 201.9 & 64.3$\pm$1.8 & 55.7 & -0.040  $\pm$  0.014 & 0.133 $\pm$  0.025 & 5.321$\pm$0.062 & 5.403$\pm$0.033 \\
    12198 & J023705.64+125406.0 & G5 & 71 & 5834 & 1.2 & 2.1 & 39.4$\pm$0.6 & 40.3 & 14.3$\pm$0.9 & 11.2 & -0.021  $\pm$  0.015 & 0.215 $\pm$  0.050 & 7.113$\pm$0.032 & 7.178$\pm$0.019 \\
    13932 & J025930.69+062022.5 & G0 & 65 & 5950 & 0.8 & 1.1 & 21.5$\pm$0.4 & 20.9 & 8.5$\pm$1.0 & 5.8 & 0.028  $\pm$  0.017 & 0.315 $\pm$  0.077 & 7.838$\pm$0.023 & 7.886$\pm$0.020 \\
    18837 & J040217.21-013757.9 & F5 & 68 & 6472 & 1.4 & 1.0 & 64.7$\pm$1.0 & 66.0 & 23.0$\pm$1.3 & 18.2 & -0.019  $\pm$  0.015 & 0.206 $\pm$  0.045 & 6.575$\pm$0.039 & 6.619$\pm$0.020 \\
    20094 & J041829.43+355926.6 & F5 & 43 & 5550 & 0.9 & 2.5 & 63.0$\pm$1.0 & 66.5 & 23.2$\pm$1.6 & 18.4 & -0.055  $\pm$  0.017 & 0.207 $\pm$  0.053 & 6.611$\pm$0.038 & 6.645$\pm$0.021 \\
    20507 & J042340.81-034444.0 & A2V & 64 & 8840 & 2.3 & 5.8 & 303.3$\pm$3.9 & 305.9 & 97.6$\pm$2.3 & 84.4 & -0.009  $\pm$  0.013 & 0.135 $\pm$  0.021 & 4.930$\pm$0.077 & 4.939$\pm$0.041 \\
    21091 & J043111.09+111439.9 & G0 & 59 & 5825 & 1.0 & 1.8 & 37.8$\pm$0.6 & 39.2 & 14.7$\pm$1.3 & 10.9 & -0.038  $\pm$  0.017 & 0.257 $\pm$  0.064 & 7.149$\pm$0.031 & 7.207$\pm$0.019 \\
    21783 & J044046.82+301728.9 & F5 & 64 & 6365 & 1.2 & 0.3 & 51.1$\pm$0.8 & 52.0 & 18.1$\pm$1.0 & 14.4 & -0.018  $\pm$  0.015 & 0.207 $\pm$  0.045 & 6.843$\pm$0.038 & 6.879$\pm$0.021 \\
    21918 & J044248.88+121233.0 & G5 & 56 & 5642 & 1.8 & 3.7 & 138.1$\pm$2.0 & 138.3 & 44.7$\pm$1.5 & 38.5 & -0.002  $\pm$  0.015 & 0.139 $\pm$  0.029 & 5.720$\pm$0.054 & 5.855$\pm$0.028 \\
    26395 & J053708.78-114632.0 & A2V & 63 & 9099 & 1.4 & 0.5 & 124.1$\pm$1.8 & 119.6 & 73.4$\pm$2.1 & 33.0 & 0.036  $\pm$  0.014 & 0.551 $\pm$  0.013 & 5.910$\pm$0.051 & 5.978$\pm$0.022 \\
    39947 & J080930.03-515033.6 & G0V & 57 & 5959 & 2.4 & 2.0 & 259.3$\pm$3.6 & 264.3 & 84.1$\pm$2.1 & 73.5 & -0.019  $\pm$  0.014 & 0.126 $\pm$  0.022 & 5.040$\pm$0.074 & 5.132$\pm$0.036 \\
    42333 & J083750.09-064824.2 & G0 & 24 & 5817 & 1.0 & 0.9 & 235.2$\pm$3.2 & 234.3 & 76.2$\pm$2.1 & 65.1 & 0.004  $\pm$  0.014 & 0.145 $\pm$  0.024 & 5.156$\pm$0.079 & 5.271$\pm$0.035 \\
    42438 & J083911.67+650116.5 & G1.5Vb & 14 & 5902 & 0.9 & 0.8 & 625.6$\pm$8.1 & 613.5 & 198.7$\pm$3.7 & 170.6 & 0.019  $\pm$  0.013 & 0.142 $\pm$  0.016 & 4.098$\pm$0.106 & 4.210$\pm$0.059 \\
    43273 & J084855.82+724034.7 & G0 & 67 & 5997 & 1.1 & 1.5 & 38.1$\pm$0.5 & 38.2 & 13.8$\pm$1.0 & 10.6 & -0.002  $\pm$  0.014 & 0.229 $\pm$  0.057 & 7.163$\pm$0.028 & 7.231$\pm$0.022 \\
    58083 & J115442.60+030837.0 & K2 & 40 & 4728 & 0.7 & 1.5 & 34.2$\pm$0.5 & 36.2 & 13.2$\pm$1.2 & 10.1 & -0.059  $\pm$  0.017 & 0.238 $\pm$  0.067 & 7.284$\pm$0.029 & 7.359$\pm$0.020 \\
    66322 & J133531.56-220128.7 & F7/F8V & 49 & 6374 & 1.4 & 1.4 & 122.0$\pm$1.7 & 125.3 & 40.3$\pm$1.3 & 34.8 & -0.028  $\pm$  0.014 & 0.137 $\pm$  0.028 & 5.892$\pm$0.053 & 5.924$\pm$0.026 \\
    67837 & J135343.46-782450.1 & G5V & 56 & 5474 & 0.8 & 3.5 & 28.4$\pm$0.4 & 29.2 & 10.3$\pm$0.7 & 8.1 & -0.029  $\pm$  0.014 & 0.214 $\pm$  0.054 & 7.485$\pm$0.025 & 7.546$\pm$0.019 \\
    70022 & J141940.92+002303.6 & A7V & 63 & 7950 & 1.7 & 0.6 & 147.5$\pm$2.0 & 152.6 & 48.9$\pm$1.6 & 42.1 & -0.035  $\pm$  0.014 & 0.138 $\pm$  0.029 & 5.680$\pm$0.061 & 5.697$\pm$0.028 \\
    72066 & J144428.29+451109.4 & F0 & 62 & 7233 & 1.6 & 0.3 & 118.1$\pm$1.5 & 118.9 & 39.1$\pm$1.2 & 32.8 & -0.007  $\pm$  0.013 & 0.160 $\pm$  0.026 & 5.930$\pm$0.051 & 5.972$\pm$0.024 \\
    73772 & J150447.01-511505.2 & G3V & 71 & 5966 & 1.1 & 0.5 & 35.9$\pm$0.6 & 36.7 & 13.1$\pm$0.9 & 10.2 & -0.022  $\pm$  0.017 & 0.221 $\pm$  0.052 & 7.233$\pm$0.030 & 7.271$\pm$0.021 \\
    78466 & J160105.03-324145.9 & G3V & 47 & 5652 & 1.1 & 1.8 & 84.6$\pm$1.2 & 86.4 & 28.7$\pm$1.3 & 24.0 & -0.021  $\pm$  0.014 & 0.162 $\pm$  0.037 & 6.332$\pm$0.046 & 6.351$\pm$0.021 \\
    85354 & J172630.24-130924.7 & K2* & 57 & 4708 & 0.8 & 0.7 & 23.0$\pm$0.4 & 23.5 & 9.4$\pm$1.1 & 6.5 & -0.020  $\pm$  0.017 & 0.303 $\pm$  0.081 & 7.752$\pm$0.024 & 7.832$\pm$0.020 \\
    92270 & J184816.42+233053.0 & F8V & 29 & 6318 & 1.2 & 0.9 & 294.5$\pm$4.1 & 312.2 & 94.9$\pm$2.4 & 86.7 & -0.060  $\pm$  0.015 & 0.086 $\pm$  0.023 & 4.940$\pm$0.069 & 4.929$\pm$0.041 \\
    100469 & J202227.53-420259.2 & A0V & 66 & 9641 & 1.7 & 2.1 & 163.9$\pm$2.3 & 176.6 & 55.4$\pm$2.0 & 48.7 & -0.078  $\pm$  0.015 & 0.121 $\pm$  0.032 & 5.550$\pm$0.066 & 5.528$\pm$0.032 \\
    110365 & J222112.66+084051.9 & G0 & 71 & 5843 & 0.9 & 1.6 & 24.2$\pm$0.4 & 24.8 & 9.6$\pm$0.9 & 6.9 & -0.024  $\pm$  0.017 & 0.282 $\pm$  0.069 & 7.656$\pm$0.023 & 7.704$\pm$0.020 \\
    115527 & J232406.43-073302.6 & G5 & 30 & 5654 & 0.9 & 1.3 & 116.4$\pm$1.5 & 120.1 & 38.9$\pm$1.4 & 33.4 & -0.032  $\pm$  0.013 & 0.140 $\pm$  0.031 & 5.939$\pm$0.056 & 5.998$\pm$0.024 \\
    117972 & J235541.67+250838.8 & G5 & 50 & 4653 & 1.4 & 4.6 & 85.6$\pm$1.3 & 87.8 & 26.0$\pm$1.1 & 24.5 & -0.026  $\pm$  0.015 & 0.057 $\pm$  0.041 & 6.418$\pm$0.045 & 6.391$\pm$0.021\\
    \enddata
    \tablecomments{\hip\ stars with detected mid-IR excesses at either $W3$ or $W4$. Unless otherwise noted, the stellar temperature and radius were obtained from photospheric model fits to the optical through 4.5$\micron$ photometry, as described in Section~3 of \p14.}
    \tablenotetext{a}{Spectral types are from the \hip\ catalog. Stars marked with asterisks have had their spectral types estimated from their $B_T-V_T$ colors using empirical color relations from \citet{Pecaut2013}.}
    \tablenotetext{b}{Parallactic distances from \hip.}
    \tablenotetext{c}{The quoted fractional excesses in $W3$ and $W4$ represent the ratios of the measured excesses and the total fluxes in these bands.  They have not been color-corrected for the filter response, although such corrections have been applied to the estimates of the fractional bolometric luminosities $f_d$ of the dust (Table~\protect \ref{tab:diskcandidates}; see Section 3 of \p14).}
    \tablenotetext{d}{Saturation corrected $W1$ and $W2$ photometry (see Section~2.4 in \p14).}
    \end{deluxetable}
    
    In most cases we used the $W4$ excess and the 3-$\sigma$ upper limits to the $W3$ excess to calculate upper limits to the blackbody dust temperatures.  In cases with significant or marginal $W3$ excesses, we calculated the actual blackbody dust temperatures. These are cases for where the $W3$ excess flux is calculated to be $>3\sigma$ below the photosphere. This is because we found that the empirically derived $W1-W3$ and $W2-W3$ photospheric colors are mostly negative (see Figures 3 of \p14). Hence, if relative to $W1$ and $W2$, the $W3$ fluxes are underestimated with respect to a Rayleigh-Jeans emission, scaling our photospheric model results in an overestimation of the model convolved $W3$ photospheric flux. 
    
    In the following section, we discuss the new excesses in the context of archival data and of the published literature to assess their reliability and, wherever possible, to elucidate the properties of the dust.

    \subsection{New Candidate Debris Disks}\label{sec:new_disks}
    
    Out of the 28 \WS\ candidate debris disks discovered since \p14, 19 are completely new detections with no previously reported excesses at any wavelength. Eighteen of these occur at $W4$, and are indicated with `Y-' in the column labeled `New?' in Table~\ref{tab:excessstats}. These are new excesses at 22\micron\ with no significant 12\micron\ excess emission.  One of the 18 new $W4$ excesses, associated with HIP~20507, is detected only as a weighted-color excess without showing any significant excess in the individual colors.

\begin{deluxetable}{lcccccccc}
\tablewidth{0pt}
\tabletypesize{\scriptsize}
\tablecaption{Debris Disk Parameters from Single-Temperature Blackbody Fits\label{tab:diskcandidates}}

\tablehead{\colhead{HIP ID} & \colhead{$T_{BB}$} & \colhead{$T_{BB_{lim}}$} & \colhead{$R_{BB}$} & \colhead{$R_{BB_{lim}}$} & \colhead{$\theta$} & \colhead{$f_d$} & \colhead{$f_{d_{lim}}$} & \colhead{Notes} \\ 
\colhead{} & \colhead{(K)} & \colhead{(K)} & \colhead{(AU)} & \colhead{(AU)} & \colhead{($\arcsec$)} & \colhead{($10^{-5}$)} & \colhead{($10^{-5}$)} & \colhead{}}

\startdata
1893 & \nodata & $<$145 & \nodata & $>$3.4 & $>$0.063 & 6.6 & $>$0.25 & b,f \\
2852 & \nodata & $<$99 & \nodata & $>$21 & $>$0.43 & 3.1 & $>$0.066 & b,f \\
12198 & \nodata & $<$185 & \nodata & $>$2.7 & $>$0.038 & 6.3 & $>$0.25 & b,f \\
13932 & 166 & $<$264 & 2.3 & $>$0.9 & 0.014--0.035 & 10 & $>$0.39 & c,f \\
18837 & \nodata & $<$197 & \nodata & $>$3.4 & $>$0.05 & 4.5 & $>$0.17 & b,f \\
20094 & 131 & \nodata & 3.9 & \nodata & 0.091 & 7.6 & $>$0.27 & a,f \\
20507 & \nodata & $<$260 & \nodata & $>$6.0 & $>$0.094 & 1.6 & $>$0.04 & b,f \\
21091 & \nodata & $<$131 & \nodata & $>$4.4 & $>$0.075 & 8.8 & $>$0.31 & b,f \\
21783 & \nodata & $<$202 & \nodata & $>$2.7 & $>$0.042 & 4.8 & $>$0.18 & b,f \\
21918 & \nodata & $<$339 & \nodata & $>$1.1 & $>$0.02 & 7.9 & $>$0.16 & b,f \\
26395 & 146 & \nodata & 13 & \nodata & 0.2 & 8.5 & \nodata & g \\
39947 & \nodata & $<$248 & \nodata & $>$3.2 & $>$0.057 & 3.9 & $>$0.12 & b,f \\
42333 & 117 & $<$344 & 5.5 & $>$0.64 & 0.027--0.23 & 5 & $>$0.15 & c,f \\
42438 & 219 & $<$432 & 1.6 & $>$0.4 & 0.028--0.11 & 4 & $>$0.14 & c,f \\
43273 & \nodata & $<$229 & \nodata & $>$1.7 & $>$0.025 & 7.1 & $>$0.24 & b,f \\
58083 & 131 & \nodata & 2.1 & \nodata & 0.053 & 15 & $>$0.53 & a,f \\
66322 & \nodata & $<$188 & \nodata & $>$3.6 & $>$0.074 & 2.8 & $>$0.11 & b,f \\
67837 & \nodata & $<$145 & \nodata & $>$2.7 & $>$0.048 & 7.8 & $>$0.3 & b,f \\
70022 & \nodata & $<$140 & \nodata & $>$13 & $>$0.2 & 1.6 & $>$0.057 & b,f \\
72066 & \nodata & $<$258 & \nodata & $>$2.9 & $>$0.046 & 3 & $>$0.089 & b,f \\
73772 & \nodata & $<$199 & \nodata & $>$2.3 & $>$0.033 & 6.3 & $>$0.24 & b,f \\
78466 & \nodata & $<$204 & \nodata & $>$2.1 & $>$0.044 & 5.1 & $>$0.19 & b,f \\
85354 & \nodata & $<$170 & \nodata & $>$1.4 & $>$0.025 & 19 & $>$0.74 & b,f \\
92270 & 131 & \nodata & 6.9 & \nodata & 0.24 & 1.9 & $>$0.067 & a,f \\
100469 & 131 & \nodata & 21 & \nodata & 0.32 & 0.88 & $>$0.027 & a,f \\
110365 & \nodata & $<$166 & \nodata & $>$2.7 & $>$0.037 & 9 & $>$0.35 & b,f \\
115527 & \nodata & $<$140 & \nodata & $>$3.3 & $>$0.11 & 4.3 & $>$0.16 & b,f \\
117972 & 367 & $>$283 & 0.31 & $<$0.87 & 0.0062 & 23 & $>$19 & d,e \\
\enddata
\tablecomments{The columns list blackbody temperatures of thermal excesses, inferred separations from the star and fractional bolometric luminosities.\\
Notes:\\
a. $W4$-only excess: The $W3$ excess flux in this case was $>3\sigma$ below the photosphere. A limiting temperature and radius for the dust cannot be determined. See detailed explanation in Section~\protect \ref{sec:results}.\\
b. $W4$-only excess: The $W3$ excess flux is formally negative and an upper limit on the excess flux is used to place a $3\sigma$ limit on the dust temperature and radius.\\
c. $W4$-only excess: Both the $W3$ and the $W4$ excesses were used to calculate a dust temperature and radius. A 3$\sigma$ upper limit on the $W3$ excess flux was used to calculate a $3\sigma$ limit on the dust temperature and radius.\\
d. $W3$-only excess: Both the $W3$ and the $W4$ excesses were used to calculate a dust temperature and radius. A 3$\sigma$ upper limit on the $W4$ excess flux was used to calculate a $3\sigma$ limit on the dust temperature and radius.\\
e. A lower limit on the fractional luminosity was calculated  for a blackbody with peak emission at $\lambda=12\mu m$ as described in Section~3 in \p14.\\
f. A lower limit on the fractional luminosity was calculated  for a blackbody with peak emission at $\lambda=22\mu m$ as described in Section~3 in \p14\\
g. Significant excesses were found both at $W3$ and $W4$. The dust parameters are calculated exactly using a blackbody for the excess.
} 
\end{deluxetable}

The remaining one of the 19 new candidate excesses, associated with HIP~117972, is significant only at $W3$, and only in the $W1-W3$ color.  It has $\Sigma_{E[W1-W3]}=2.73$: just above the $\Sigma_{E[W1-W3]_{98}}=2.66$ 
confidence level threshold.  It is not confirmed as a weighted-color excess at $W3$ because the weighted $W3$ excess confidence threshold is higher: at $\Sigma_{\overline{E[W3]_{98}}}=3.28$.  Given our adoption of a lower confidence level (98\%) for detecting $W3$ excesses, it is possible that the excess from HIP~117972 may be spurious.  Nonetheless, the star does show a marginal excess also in the $W1-W4$ and $W2-W4$ colors.  The combined evidence for faint $W3$ and $W4$ excesses suggests that they may be real, and that HIP~117972 may host a warm zodiacal dust-like debris disk.  A joint SED fit to the shorter-wavelength and WISE photometry indicates a $\sim$531~K dust excess (Figure~\ref{fig:SEDs}, bottom left panel) at $f_d=1.92\times10^{-4}$ of the stellar bolometric luminosity (Table~\ref{tab:diskcandidates}).

	\subsubsection{New Disk Candidates with Archival IR Observations} \label{sec:newdisk_archival}

\begin{figure}[htb!]
    \centering
    \begin{tabular}{cc}
    \includegraphics[scale=0.4]{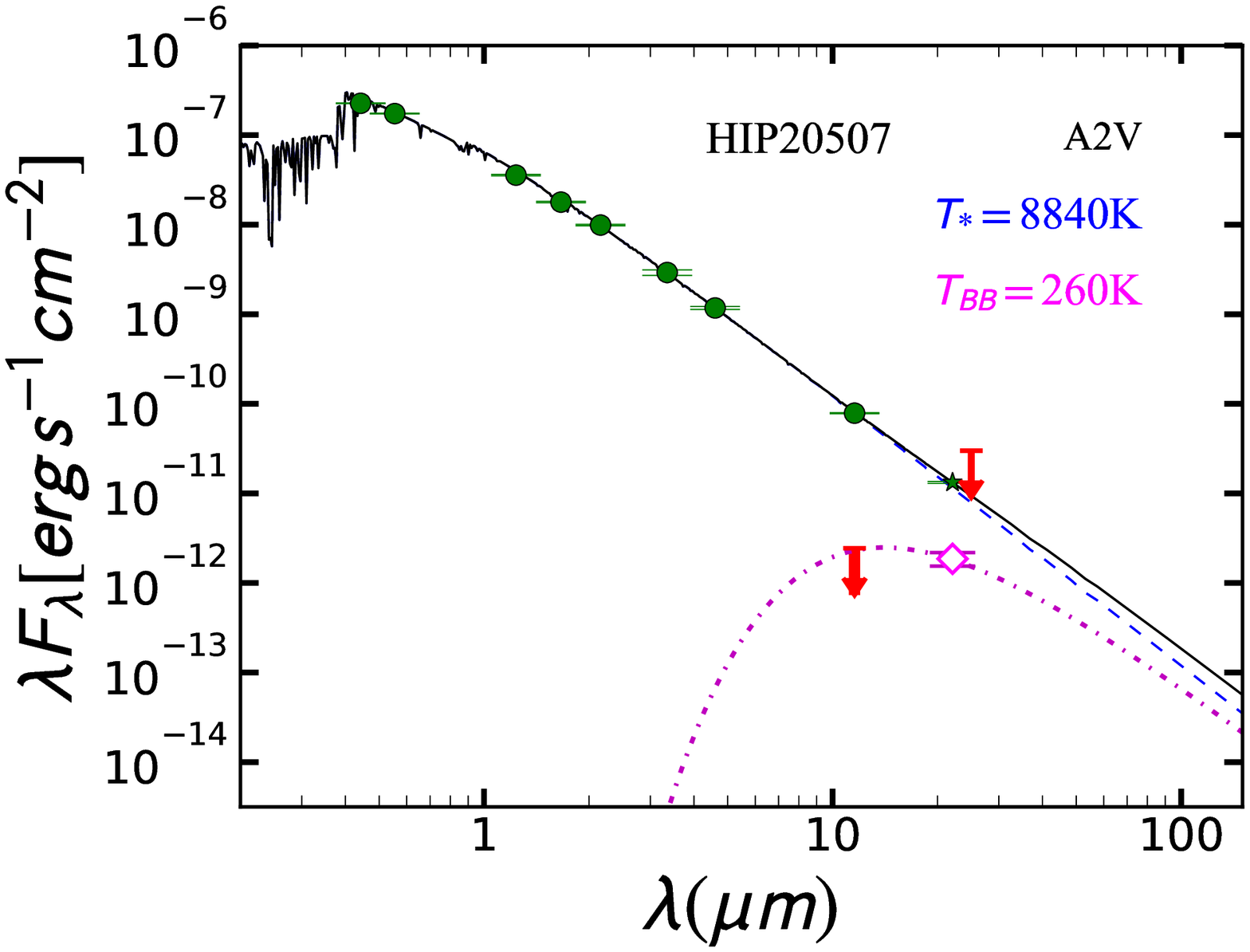} &
    \includegraphics[scale=0.4]{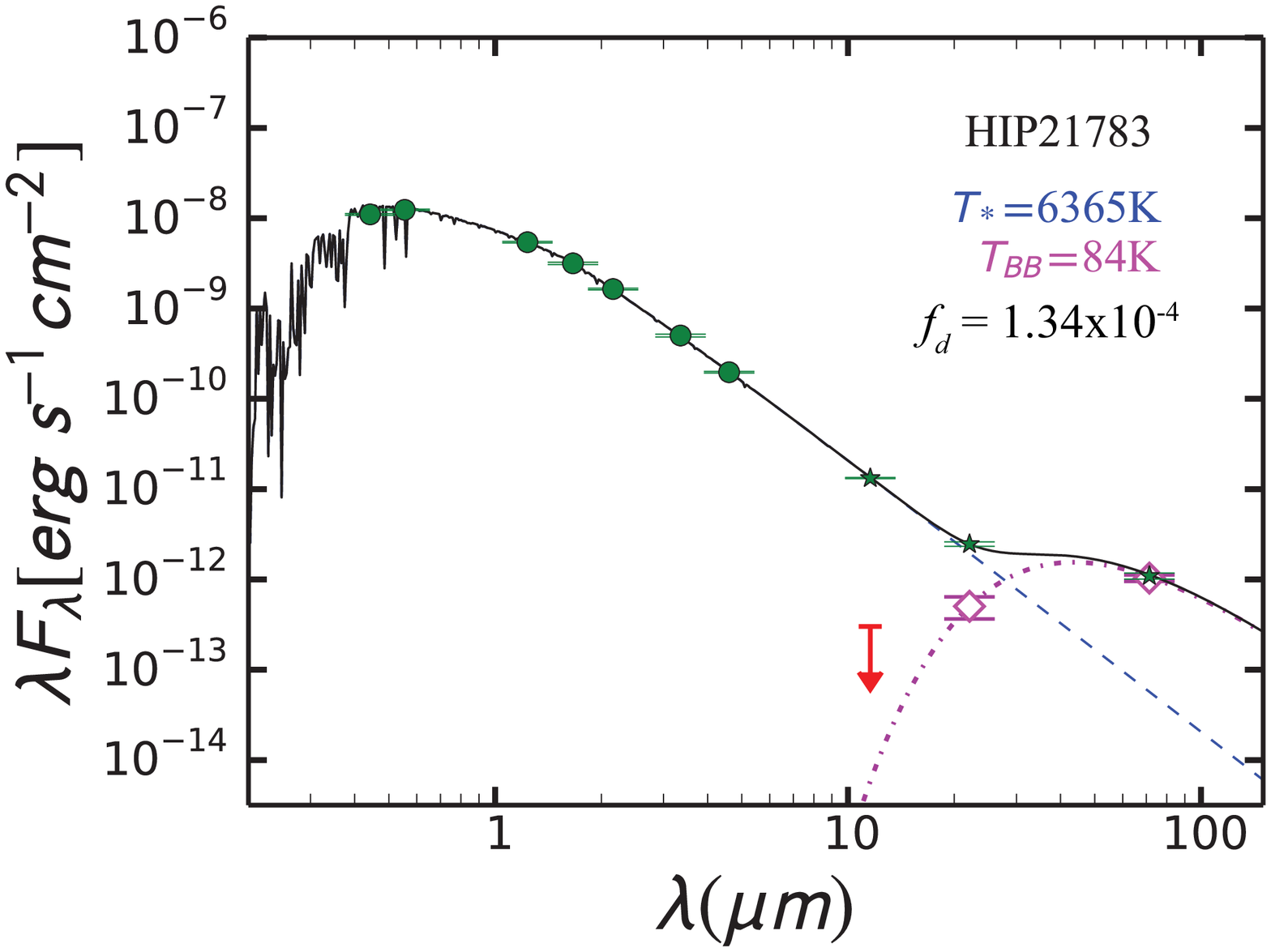} \\
    \includegraphics[scale=0.4]{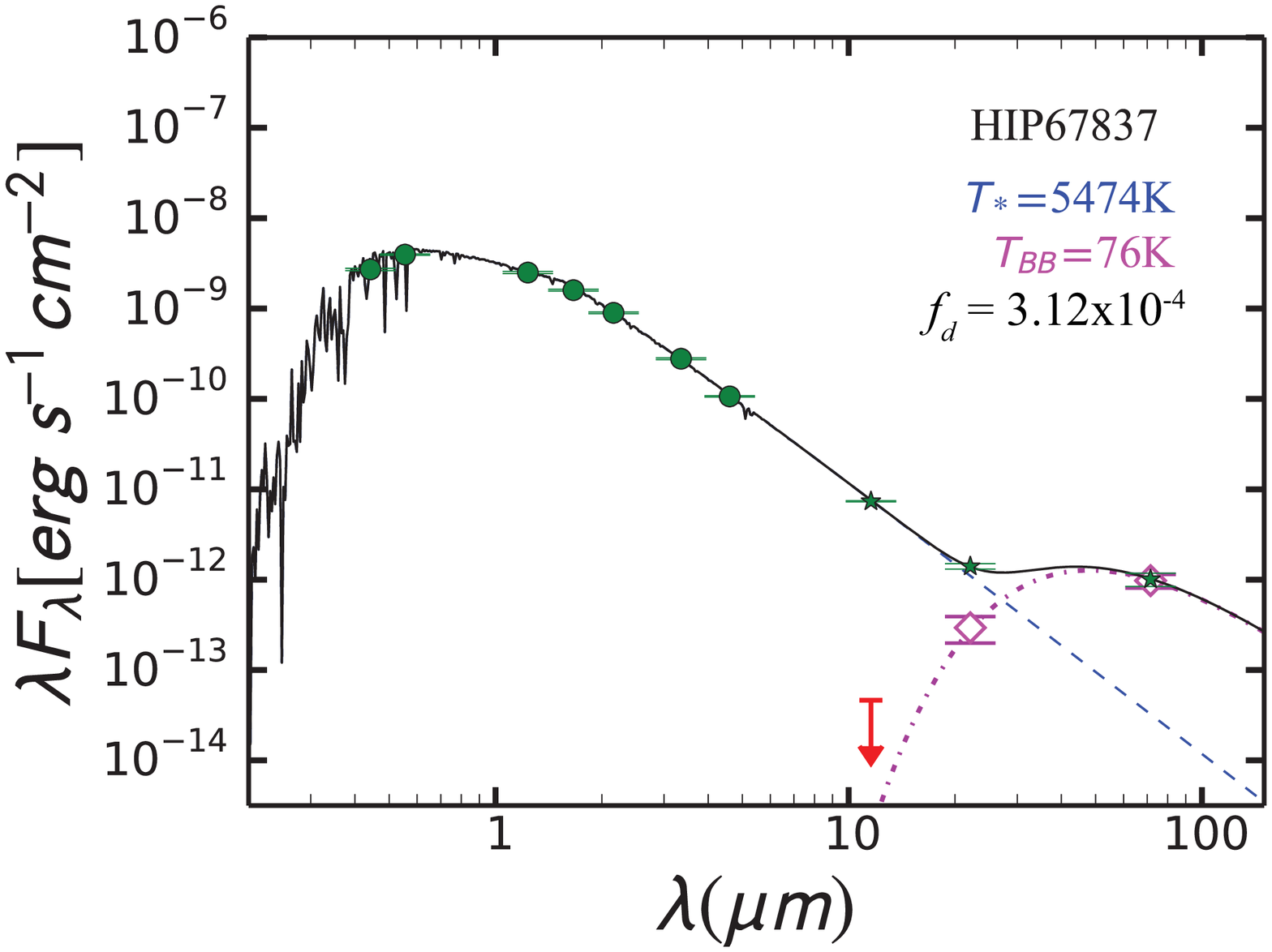}  &
    \includegraphics[scale=0.4]{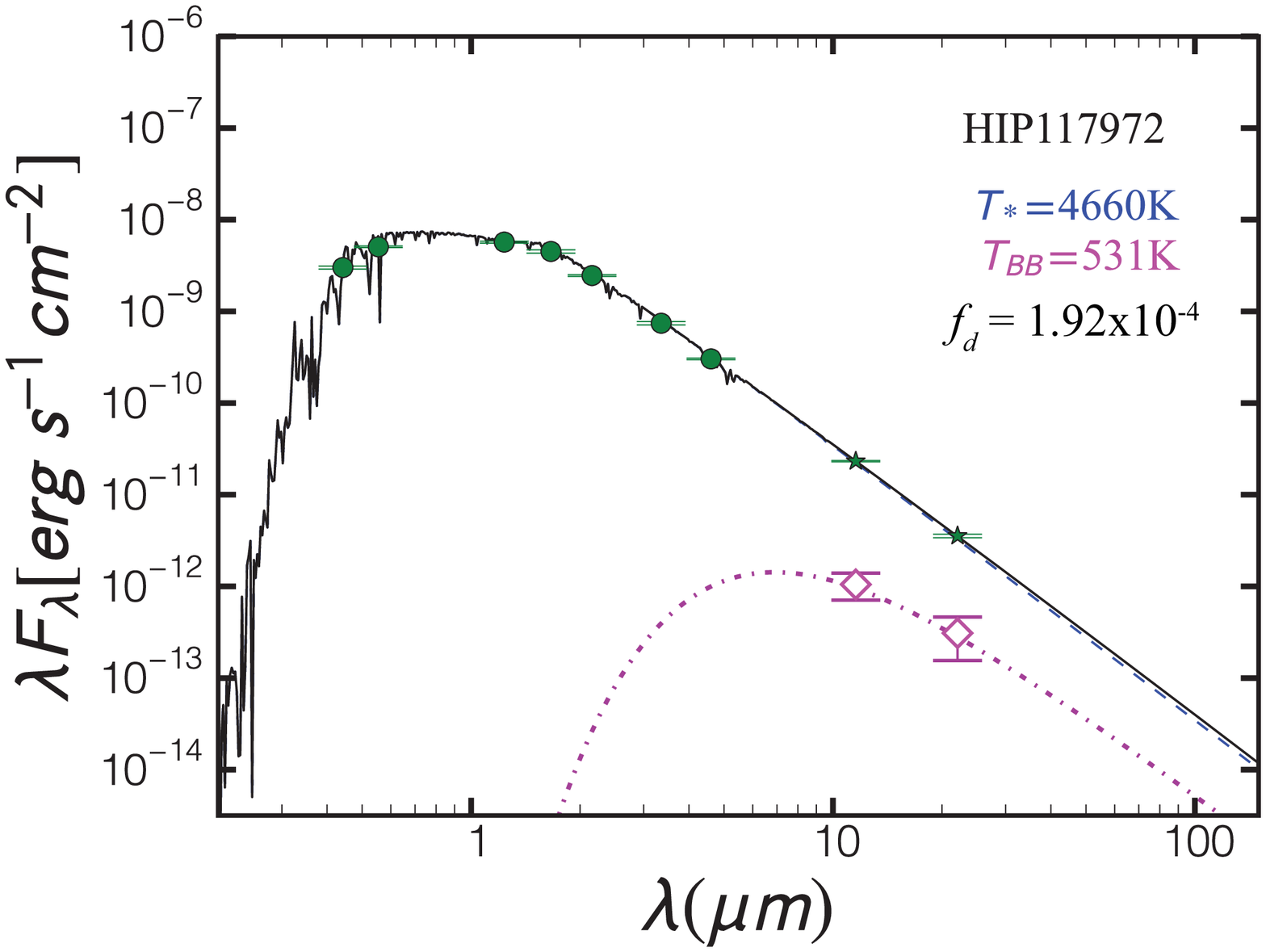}
    
    \end{tabular}
    \caption{Example SEDs representative of newly detected excesses from this study. The blue dashed lines correspond to the fitted NextGen photosphere models to photometry from the \hip\ catalog (Johnson $B, V$), \mass\ catalog ($J, H, K_s$), and \WS\ All-Sky Catalog  ($W1, W2$)  photometry. For HIP~20507, we also fit the photosphere using $W3$ --- as indicated by the green circles. After fitting, the photosphere was further scaled to the weighted average of the $W1$ and $W2$ fluxes to take advantage of the synchronicity and uniform calibration of all \WS\ photometry. The $W1$ and $W2$ photometry were corrected for saturation following \p14. $W3$ and $W4$ All-Sky photometry are green stars at 12 and 22\micron\ in each plot. We fit blackbody curves (magenta dashed-dot curves) to excess fluxes (open magenta diamonds) and 3$\sigma$ upper limits (red arrows) red-ward of $W3$. The combined photosphere and excess emission for each star is plotted as solid black line. HIP~21783 and HIP~67837 are new $W4$ excesses we identified from the significance of their $W2-W4$ and $W3-W4$ color, respectively. We also use archival {\it Spitzer}/MIPS 70\micron\ and {\it Herschel}/PACS 70\micron\ fluxes to further constrain the dust temperature fits for HIP~21783 and HIP~67837, respectively. The \spitzer\ and {\it Herschel} fluxes were obtained as described in Section~\ref{sec:newdisk_archival}. In addition, HIP~117972 is a new $W3$-only excess which we identified from the significance of its $W1-W3$ color, while HIP~20507 is a new weighted $W4$ excess. The upper-limit \iras\ 25\micron\ flux is plotted, although it does not provide any useful constraints.\label{fig:SEDs}}
    \end{figure}

While none of the stars with new candidate excess detections discussed here have been previously identified as debris disk hosts in the literature, perusal of archival observations from \iras\, \spitzer, {\it Herschel}, and {\it AKARI} reveals data for HIP~910, HIP~20507, HIP~21783, and HIP~67837. HIP~20507 has only \iras\ data at 25\micron, though the detection is too noisy to place useful constraints and hence we do not include it in our SED fit (Figure~\ref{fig:SEDs}, bottom right panel). We discuss the other three candidate excesses with archival observations below, noting that the small HIP~910 $W4$ excess found by us is likely spurious. Hence, our total number of new \WS\ excesses is in fact 19.

\paragraph{HIP 910.}
Among the four stars for which archival mid-IR data exist, only HIP~910 has been discussed in the debris disk literature, where it has received considerable scrutiny as a nearby \citep[19~pc;][]{VanLeeuwen2007} near-solar analog \citep[F8V;][]{Gray2006}.  Independent analyses of {\it Spitzer}/IRS low-resolution spectra \citep{Beichman2006}, {\it Spitzer}/MIPS 24\micron\ and 70\micron\ photometry \citep{Trilling2008}, and {\it Herschel}/PACS 100\micron\ and 160\micron\ photometry \citep{Eiroa2013} all conclude that HIP~910 does not possess an excess. We find that HIP~910 has small but significant $W2-W4$ ($0.19\pm0.06$~mag) and $W2-W3$ ($0.15\pm0.04$~mag) excesses above the photosphere.  As such, HIP~910 would be a candidate for having a zodiacal dust debris disk analog.  The inferred 19\% excess at $W4$ would have only been $\sim$2$\sigma$ significant in the MIPS24 observations of \citet{Trilling2008}, hence the non-confirmation in MIPS is not surprising.  However, the 15\%--19\% excess over 10--30\micron\ would have been detected at $\sim$10$\sigma$ significance in the {\it Spitzer}/IRS analysis of \citet{Beichman2006}.  Their low-resolution {\it Spitzer}/IRS observations cover a wide wavelength range, 6--38\micron, and have superior sensitivity to faint excesses compared to our \WS\ photometric analysis: because of the better stellar photospheric estimation that is attainable with a larger number of independent short-wavelength data points.  Given the lack of confirmation from the {\it Spitzer}/IRS observations, we conclude that the candidate $W4$ excess from HIP~910 is probably spurious: likely the result of a $W2$ measurement that is $>$3$\sigma$ below the photosphere.  HIP~910 may be representative of the very few ($\lesssim$2) $W4$ false-positive excesses expected beyond our 99.5\% FDR threshold.

HIP~910 is the only newly-identified excess candidate in the present study for which published mid-IR observations exist.  Because it is also unique in that it is not confirmed as a debris disk in the more sensitive {\it Spitzer}/IRS data, this raises the question whether some of our other candidates discussed here and in \p14 may also be spurious.  To determine whether the non-confirmation of \WS\ excesses from {\it Spitzer}/IRS observations is a common occurrence for any of our reported excesses, we searched the recent literature for all of the new excess stars discovered in \p14.
Nineteen of these have had {\it Spitzer}/IRS observations published since, all in \citet{Chen2014}. All are confirmed to have {\it Spitzer}/IRS excesses.\footnote{After the publication of \p14 we further recognized that some of the excesses that we had reported as new had already been identified as candidate debris disks from {\it Spitzer}/IRS spectra by \citet{Ballering2013}.  There are 14 such excesses: a subsample of the 19 new \p14 $W4$ excesses confirmed in \citet{Chen2014}.} Hence, we can conclude that the non-confirmation of HIP~910 is not typical of our \WS\ excess detections, and that the remaining 19 new candidate debris disks reported here and the 104 new candidates in \p14 remain viable.

\paragraph{HIP 21783.}
This star is serendipitously included in a single MIPS 70\micron\ pointing in {\it Spitzer} program GO~54777 (PI: T.~Bourke).  We measure a flux of $26\pm2$~mJy from $r=$16\arcsec\ aperture photometry on the post-basic calibrated data (PBCD) images, after an aperture correction factor of 2.04.\footnote{Following Table~4.14 of the MIPS Instrument Handbook v.\ 3.0; \url{http://irsa.ipac.caltech.edu/data/SPITZER/docs/mips/mipsinstrumenthandbook/}}  The MIPS70 measurement confirms the presence of a thermal excess.  A fit to the optical--IR SED (Figure~\ref{fig:SEDs}, top left panel) reveals that the associated circumstellar dust has a temperature of 84~K and a fractional luminosity of $f_d=1.34\times10^{-4}$.

\paragraph{HIP 67837.}
HIP~67837 is included in a {\it Herschel}/PACS 70\micron\ and 160\micron\ Open Time program (PI: D.~Padgett).  Its 70\micron\ flux is $24\pm4$~mJy, where we have performed $r=5$\arcsec\ aperture photometry on the Level 2.5-processed images, and applied an aperture correction factor of $1/0.577=1.733$ \citep[following Table 2 of][]{Balog2013}. The PACS 70\micron\ measurement confirms the thermal excess (Figure~\ref{fig:SEDs}, top right panel).  The star is not detected at 160~\micron.  The inferred dust temperature is 76~K and the fractional dust luminosity is $f_d=3.12\times10^{-4}$.

	\subsubsection{New Disk Candidates in Binary Systems}

Two of our new excess stars, HIP~2852 and HIP~70022, have M-dwarf companions \citep{DeRosa2014}.  This may be a cause for concern, as these companions might be responsible for the $W4$ excesses from these two stars. HIP~2852 has a physical 0.30$M_\sun$ companion, which corresponds to an M3/4 spectral type, at a separation of 0.93\arcsec\ $\pm$ 0.01\arcsec\ ($45.6\pm0.49$~au). HIP~70022 has a 0.18~$M_\sun$ (M5/6) companion that is also likely physical \citep{DeRosa2014}, separated by 1.84\arcsec\ (116~au) from the central star. Given $\Delta K_s\geq5$ mag contrasts between the primaries and the companions in both cases, the fluxes from the respective M-dwarf companions are not enough to produce the observed 13\%--16\% $W4$ excesses.  Therefore, we conclude that both stars possess real mid-IR excesses that are likely associated with debris disks. After factoring the companion separation for both of these stars, the dust in each system is expected to be circumprimary and not circumbinary.
    
        \subsection{Confirmation of Previously Reported 22\micron\ Faint Debris Disks}\label{sec:confirmed_disks}

    In Section~\ref{sec:newdisk_archival}, we discussed all 19 new debris disks reported in the present work. We now discuss the nine additional debris disk excesses that have been published by other teams and that we recover here, but that were not identified in \p14. Amongst them, is HIP~26395, a star for which we report a new small $W3$ excess. We had previously identified a $W4$-excess for HIP~26395 in \p14.  
    
    Five of the $W4$ excesses have been independently reported as such from \WS: four by \citet[][; HIP~12198, HIP~21091, HIP~78466 and HIP~115527]{Vican2014} and one by \citet[][;HIP~92270]{Mizusawa2012}.  We determine upper limits on the dust temperatures in these systems (Table~\ref{tab:diskcandidates}) as we have done for the newly reported debris disks (Section~\ref{sec:results}) and in \p14.  Our dust temperature limits are consistent with, albeit generally more stringent (131--203~K) than reported in \citet{Vican2014} for the four stars in common.  We use the individual 3-$\sigma$ upper limits on the $W3$ excess fluxes, rather than assume a uniform 200~K dust temperature upper limit based on the lack of $W3$ excesses.  No dust temperature information is given by \citet{Mizusawa2012} for the fifth star.

    Three of the $W4$ excess hosts (HIP~42333, HIP~42438 and HIP~100469) have published mid- and far-IR excess detections from {\it Spitzer}.  The longer-wavelength detections affirm the existence of debris disks around these stars, and provide greater constraints on the dust properties in these systems. \citet{Plavchan2009} reported MIPS 24\micron\ and 70\micron\ excess detections for HIP~42333 and calculated the dust temperature of the excess to be $T<91$~K. Our estimates of the blackbody dust temperature solely from the $W4$ excesses and the $W3$ 3-$\sigma$ upper limit yield a hotter, yet consistent result ($T_{BB}<344$~K). HIP~42438 and HIP~100469 are both known to have excesses between 8--30\micron\ from {\it Spitzer}/IRS and at 70\micron\  from {\it Spitzer}/MIPS. \citet{Chen2014} report multi-temperature debris disks for both stars, with $\sim$70--80~K cold dust components and $<$499~K warm dust components. Our single-population dust temperature estimates from $W3$ and $W4$ are consistent: $T_{BB}<432$~K for HIP~42438 and $T_{BB}=131$~K for HIP~100469 (for which we measure a significant excess also at $W3$).

    Finally, HIP~26395 was already included in \p14 as a $W4$ excess, and is known to harbor cold dust with 70$\micron$ emission \citep{Ballering2013}. Here, we report the additional detection of a weighted $W3$ excess. \citet{Chen2014} independently report a 10--30\micron\ excess seen in \spitzer/IRS data. \citeauthor{Chen2014} find that HIP~26395 has a multi-temperature debris disk, similar to those around HIP~42438 and HIP~100469: a cold component at T=94~K and a hot component at T=399~K. Again, our single-population dust temperature (146~K) is consistent with the two-population dust model of \citet{Chen2014}. Notably, our detection of the weighted $W3$ excess shows that our improved technique can detect as faint a population of excesses as is detectable by \spitzer/IRS thanks to our increased precision in determining the level of the photosphere.

        \subsection{Unconfirmed \WS\ 22\micron\ Excess Candidates from the Literature}\label{sec:unconfirmed_disks}
        
Our study is constrained only to \WS\ excesses from B9--K main sequence \hip\ stars within 75~pc and outside of the galactic plane.  We compare our findings to searches for \WS\ debris disks within this volume.  The main comparison studies are those of \citet{McDonald2012, Mizusawa2012, Wu2013, Cruz-SaenzdeMiera2014, Vican2014}, and most recently, \citet{Cotten2016}.  

Similarly to our approach, \citet{Mizusawa2012, Wu2013}, and \citet{Cruz-SaenzdeMiera2014} used \WS\ colors, at least in part, to seek mid-IR excesses from debris disks.  As already discussed in PMH14, we reliably recover all of the excesses reported in \citet{Wu2013} and \citet{Cruz-SaenzdeMiera2014} that pass our strict photometric quality selection criteria.  This is also largely the case for the \citet{Mizusawa2012} work, although we do not recover five of their 22 candidates because they are either outside of our search region (HIP~55897 being in the galactic plane) or suffer potential contamination:
from a close binary companion (HIP~88399), from saturation in the three shortest-wavelength \WS\ bands (HIP~61174), from other sources based on their \WS\ confusion flags (HIP~18859 and HIP~100800), or as inferred from discrepant photometry between the reported \WS\ values and the averaged single-frame measurements  \citep[HIP~18859; see Section 2.3 of][]{Patel2014}.

The set of studies by \citet{McDonald2012, Vican2014} and \citet{Cotten2016} follow a different excess search approach, comparing stellar photospheric models to optical-through-infrared SEDs that incorporate photometry from multiple instruments and epochs.  As we discussed in \p14\ and in Section~\ref{sec:intro},
this method is vulnerable to systematics induced by differences in photometric calibration among filter systems and by stellar variability. The presence of systematics is evident from the fact that (model plus) SED-based searches result in non-negligible numbers of large ``negative'' excesses, to the tune of $-5\sigma$ to $-10\sigma$. Consequently, the reliability of positive outliers at comparable numbers of standard deviations---which would be considered candidate excesses---is diminished.

Our \WS-only color-based search overcomes these systematic issues. Because we only use the measured \WS\ colors we circumvent any instrument-to-instrument and epoch-to-epoch systematics. In addition, by empirically calibrating the photospheric colors of stars in \WS, we have removed the spectral response dependence in estimating the stellar photosphere.  This latter point is particularly important as the published \WS\ filter profiles carry a residual color term depending on the slope of the mid-IR SED \citep[e.g.,][]{Brown2014}. 

We do not recover substantial fractions of the excesses reported in SED-based searches: e.g., 41 of the 81 excesses in \citet{Vican2014} that pass our selection criteria.  In some cases the $Wi-W4$ (where $i<3$) colors are in fact significantly negative (\p14), meaning that the apparent excesses are not confirmed in \WS\ data alone, and may thus be the result of the systematic uncertainties in the \WS\ photometric zero points \citep{Wright2010} or of stellar variability between the \WS\ and prior photometric epochs.  At the same time, it is not surprising that with our presently more aggressive color-excess detection thresholds (Section~\ref{sec:improved_detection}) relative to \p14, we now recover some additional candidate excesses (Section\ref{sec:confirmed_disks}) reported by \citet{Vican2014}.  A comparison to the much more comprehensive Tycho-2-based \WS\ study of \citet{Cotten2016} is forthcoming.

\section{Discussion: Single- vs. Weighted-Color Excess Searches}\label{sec:discussion}

 We have presented an improved set of procedures for detecting IR excesses in individual \WS\ colors (Section~\ref{sec:improved_detection}), and also an approach to combining the individual colors and producing a weighted-color excess metric at $W3$ or $W4$ (Section~\ref{sec:metric}).  Here we compare the two methods.
 For consistency, we perform the comparison only over the sample of stars with valid \WS\ photometry in all four bands. 
 
 The Venn diagrams in Figure~\ref{fig:Venn} show the correspondence between the single- and weighted-color excess detections in this sample. The weighted excess metrics confirm all five of the single-color $W3$ excesses, and 165/175 (94.3\%) of the single-color $W4$ excesses from \p14 and from Section~\ref{sec:improved_detection}.  Perhaps surprisingly, we find only two new excesses in the weighted-color selections: one at $W3$ and one at $W4$.
 
\begin{figure}
    \centering
    \includegraphics[scale=.6]{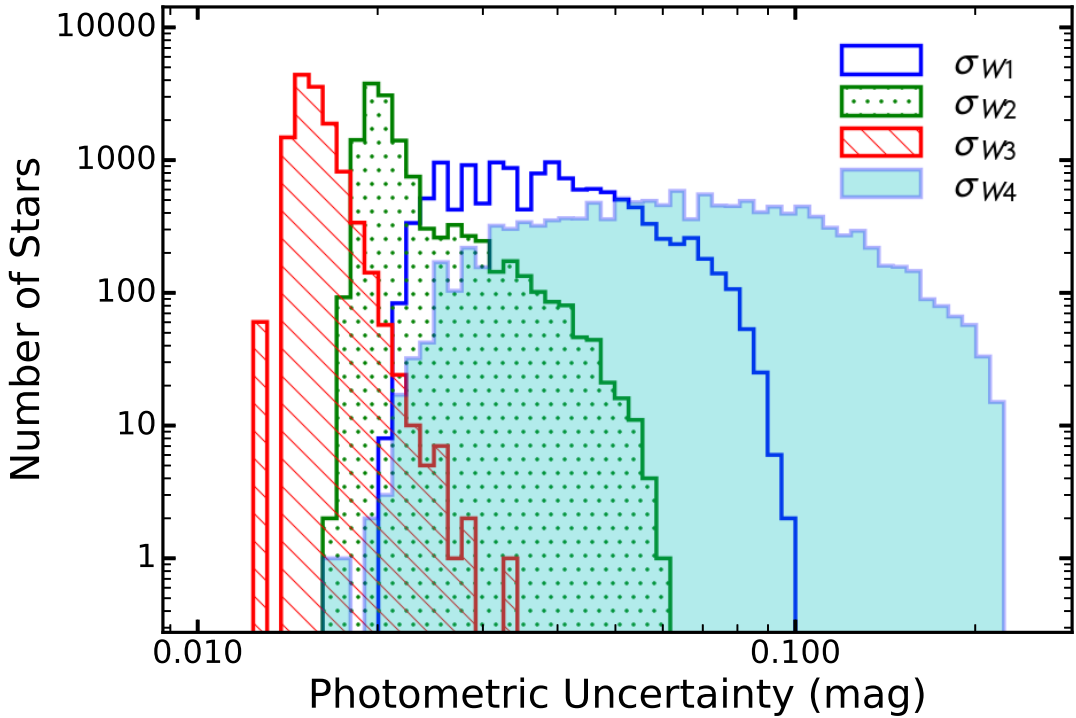}
    \caption{Distributions of photometric uncertainties for all four \WS\ bands for the 12654 stars in the weighted $W4$ parent sample, including stars with saturated and then corrected $W1$ and $W2$ photometry. The large spread in $\sigma_{W4}$ is expected because of the lower absolute flux levels in $W4$. It is evident that the mean $\sigma_{W1}$ is larger than the means of $\sigma_{W2}$ or $\sigma_{W3}$.   The $W2-W3$ color is thus in principle most sensitive to small amounts of excess, although in practice most of the detected excesses come from $W3-W4$. \label{fig:wise_errors}}
    \end{figure}

   Our initial expectation was that by averaging down the photometric uncertainties, a weighted-color excess search might have been able to produce significant detections of previously marginal single-color excesses. In reality however, all of the individual color components in our weighted-color excess measure are correlated through their common use of the same longer-wavelength filter.  Thus, the three individual $Wi-W4$ colors are correlated, and do not give independent assessments of the presence of a $W4$ excess. Consequently, the averaging in the weighted-color excess combination does not substantially improve our sensitivity.  Moreover, a consideration of the \WS\ photometric uncertainty distributions (Figure~\ref{fig:wise_errors}) shows that the $W4$ photometric errors dominate. As a result of the large $W4$ photometric errors, combining the individual $Wi-W4$ colors only marginally improves the accuracy of the $W4$ excess measurement. The weighted-color excess metric does produce higher-fidelity excesses, but only slightly so.
   
  Conversely, if a star's \WS\ single-color excess is not confirmed by the weighted-color excess metric, then the single-color excess might be considered suspect. That is, the ten stars that are not detected in our weighted $W4$ excess search (Figure~\ref{fig:Venn}b), might be false detections. Nonetheless, there are two reasons for which a star may not have a weighted $W4$ excess but may still be a bona-fide debris disk detection from a single-color excess.

\begin{figure}
    \centering
    \includegraphics[scale=0.4]{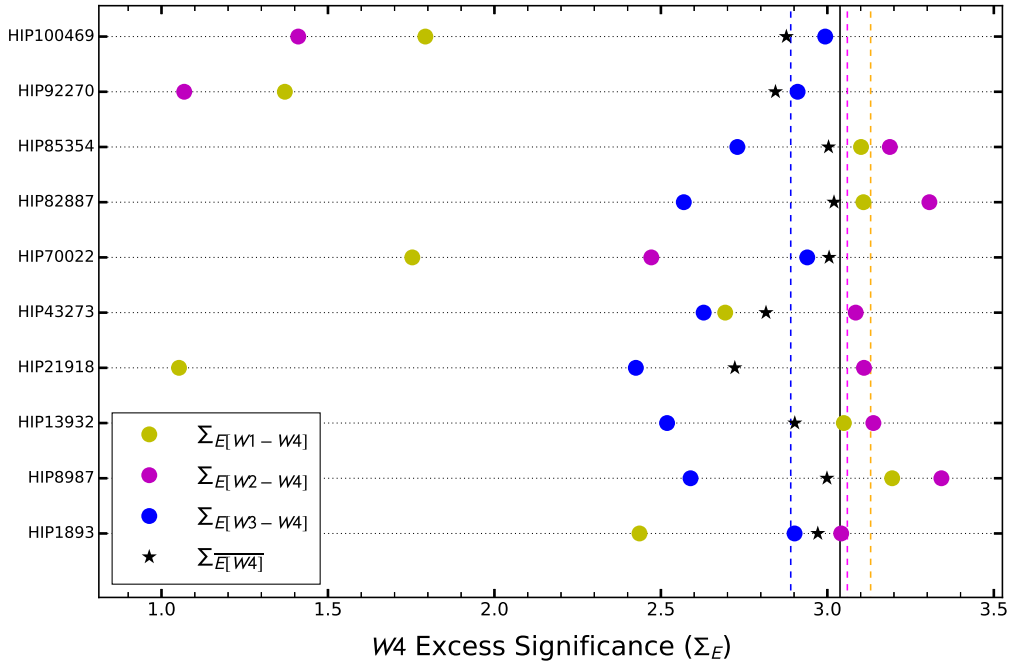}
    \caption{The excess significances for the ten stars with single-color $W4$ excesses in \p14 that were not recovered with the weighted $W4$ excess metric in this study (see Figure~\ref{fig:Venn}b). Each vertical colored line corresponds to the current 99.5\% detection threshold for each color listed in the legend. We see that the weighted $W4$ excess threshold ($\Sigma_{\overline{E[W4]}}$) effectively averages the individual single-color detection thresholds.  The stars that are not confirmed in the weighted-color selection possess significant single-color excesses in only one or two colors.\label{fig:unrecovered_10}}
    \end{figure}

   The first is that the presence of a small but positive $W3$ excess can decrease the overall significance $\Sigma_{\overline{E[W4]}}$ of the $W4$ three-color-weighted excess. Six out of the ten unrecovered stars in the weighted $W4$ search have small but positive $W1-W3$ or $W2-W3$ excesses (HIP~8987, HIP~13932, HIP~21918, HIP~43273, HIP~82887, and HIP~85354). In an attempt to potentially increase the number of new detections, we then ran a two-color weighted search by excluding the $W3-W4$ color and only using $W1-W4$ and $W2-W4$ in the weighted-color excess metric (Equation~\ref{eq:combined_significance}). However, the two-color weighted $W4$ excess search did not bear any new fruit; it produced just as many new stars when compared to the set of single-color detections as the three-color weighted search had produced. We attribute the lack of an increase in detections from the two-color weighted search to the fact that the $W3$ photometric errors are on average smaller than at $W1$ and $W2$ (Figure~\ref{fig:wise_errors}). That is, the elimination of $W3-W4$ from the weighted-color excess calculation removes a slight bias against detecting $W4$ excesses by eliminating marginally significant $W3$ excesses.  However, any gains are offset by the greater uncertainty in the $W1$ and $W2$ photometry. That is, by excluding $W3-W4$ we are excluding a large fraction of the ``excess signal,'' and leaving more of the noise (Figure~\ref{fig:unrecovered_10}).

    The fact that the $W3$ photometric errors are on average the smallest indicates that some bona-fide faint $W3-W4$ excesses may not be confirmed in $W1-W4$ and $W2-W4$, and even in the weighted $W4$ excess.  This is the second reason for which some of the single-color candidate $W3-W4$ excesses probably reveal real debris disks, even if they are not confirmed in the weighted $W4$ analysis. Such is the case for the remaining four of the ten single-color excess stars that are not recovered by the weighted-color excess metric: HIP~1893, HIP~70022, HIP~92270, and HIP~100469.   All of these are $W3-W4$-only single-color excess detections and have much larger photometric uncertainties in $W1$ and $W2$ than in $W3$: not surprising as all four stars are saturated in $W1$ and $W2$.  Even though we correct the saturated photometry of these stars, the resulting photometric uncertainties will always be larger than those of unsaturated stars.
    
\section{Conclusion}\label{sec:conclusion}

 We have presented a series of techniques that improve the ability to detect and verify the existence of \WS\ mid-IR excesses from debris disks around main sequence stars. First, we have implemented an improved assessment of the confidence threshold beyond which stars with IR excesses can be identified based on their \WS\ colors.  This has revealed 18 new potential debris disks around main-sequence \hip\ stars within 75~pc.

Second, we have presented a method that uses an optimally-weighted average of multiple \WS\ colors to identify $W3$ and $W4$ excesses, in an attempt to attain greater accuracy compared to using individual \WS\ colors. While the color weighting approach has the potential to identify fainter IR excesses, most of the excesses are expressed only at $W4$: the band with the largest $W4$ photometric uncertainties. Hence, we are unable to uncover a substantial new population of debris disks, and add only two new detections. For one of these, HIP~26395, we detected a weighted-$W3$ excess on top of the $W4$ detection found in \p14. However this star was already known as a debris disk host from previously published longer-wavelength observations. The second, HIP~20507, is the only new debris disk candidate we detected from its weighted-$W4$ excess.

    Finally, we implement an astrometric technique to discern bona-fide IR excess sources from ones that are contaminated by blends from unrelated nearby point or extended sources. We use the original unsmoothed \WS\ images available through the {\sc unWISE} service to assess the positions of the stellar centroids between $W3$ and $W4$, and between $W4$ measurements with two different aperture sizes.  We reject eleven candidate excesses with this approach, four of which had been reported in the previous literature as debris disk candidates. HIP~68593 and HIP~95793 have well established excess detections \citep[e.g.,][,respectively]{Carpenter2009,Draper2016}, while HIP~60074 has a spatially resolved cold dust disk \citep{Ardila2004}. HIP~19796 also has a \spitzer/MIPS identified excess $K_S$-[24] = 0.09~mags \citep{Stauffer2010, Urban2012}. However, given this star's relatively small excess and that we identified it as a an astrometric rejection, we feel the existence of its debris disk may be questionable. As we have stated previously, the rejection of any debris disk candidate using our astrometric technique, though it may indicate the presence of a blended background source, does not necessarily discount the existence of a circumstellar debris disk. Although we do not eliminate visual checks of the \WSAC\ images after excess identification, the automated assessment of the stellar centroid offsets provides a sensitive and objective metric to assess contamination.

    Overall, the use of a weighted-color excess combination of \WS\ colors improves the reliability of candidate IR excess detections from individual \WS\ colors at the cost of potentially overlooking a remaining small population of faint W4 excesses. Even though the fraction of debris disk-bearing stars within 75~pc does not change significantly from the findings in our previous study, the verification through weighted colors and the positional checks using higher angular resolution images provide confidence that the 19 new disks discovered here are real, and not spurious or contaminated. Thus, combined with the \p14 results, we find a total of 9 $W3$ and 229 significant $W4$ excesses from $<$75~pc \hip\ stars in \WS. As of the current study, 107 of these represent previously unreported 10--30\micron\ excesses, 101 of which represent entirely new debris disk detections within 75~pc.  This expands the 75~pc debris disk sample by 22\% around \hip\ main sequence stars and by 20\% overall (including non-main sequence and non-\hip\ stars).

\acknowledgments We thank Dustin Lang for help with downloading images for our entire sample from the {\sc unWISE} image service. We would like to acknowledge assistance from Melissa Louie who provided suggestions to improve figure aesthetics. This publication makes use of data products from the Wide-field Infrared Survey Explorer, which is a joint project of the University of California, Los Angeles, and the Jet Propulsion Laboratory/California Institute of Technology, funded by the National Aeronautics and Space Administration. We also use data products from the Two Micron All Sky Survey, which is a joint project of the University of Massachusetts and the Infrared Processing and Analysis Center/California Institute of Technology, funded by the National Aeronautics and Space Administration and the National Science Foundation. This research has also made use of the SIMBAD database, operated at CDS, Strasbourg, France. This research has made use of the Washington Double Star Catalog maintained at the U.S. Naval Observatory. Most of the figures in this work were created using Matplotlib, a Python graphics environment \citep{Hunter2007}. This research also made use of APLpy, an open-source plotting package for Python hosted at \url{http://aplpy.github.com} \citep{Robitaille2012}. This work is partially supported by NASA Origins of Solar Systems through subcontract No. 1467483 to Dr. Stanimir Metchev at Stony Brook University, and by an NSERC Discovery award to Dr. Stanimir Metchev at the University of Western Ontario.

\appendix
\section{The Weighted-Color Excess Metric}\label{sec:appendix}

We present the full derivation of $\Sigma_{\overline{E[Wj]}}$ for a star at a \WS\ mid-IR band $Wj$, where $j=3\mbox{ or }4$. Starting with Equation~\ref{eq:excess_1}, we arrive at a general form for the weighted-color excess by adding the individual color excess terms, and multiplying by weights $a_i$

\begin{eqnarray}
\overline{E[Wj]} &=&  \sum_{i=1}^{j-1} a_i E[Wi-Wj]\label{eq:gen_weighted_excess}\\
                 &=& \sum_{i=1}^{j-1} a_i\left(Wi-Wj-W_{ij}(B_T-V_T)\right)\label{eq:gen_weighted_excess2}.
\end{eqnarray}

\noindent The weights $a_i$ are normalized and are unknown:

\begin{equation}\label{eq:sum_weights}
\sum_{i=1}^{j-1} a_i \equiv 1. 
\end{equation}

\noindent Our general form for the S/N of the weighted average of the excess at $Wj$ is calculated by dividing equation~\ref{eq:gen_weighted_excess} by the uncertainty in the weighted average, $\sigma_{\overline{E[Wj]}}$. The uncertainty is defined as the quadrature sum of each entry of the Jacobian matrix of $\overline{E[Wj]}$ weighted by its respective uncertainty. The variance of the weighted average is

\begin{equation}\label{eq:wtavgExcessUnc}
      \sigma_{\overline{E[Wj]}}^2 = \sum_{\alpha} \sigma_{\alpha}^2 \left(\frac{\partial \overline{E[Wj]}}{\partial \alpha}\right)^2 + O\left(\sigma_{Wi,Wij} \right) + O(\sigma_{Wi,Wj}),
\end{equation}

\noindent where $\alpha \in \{Wi, Wj, Wij(B_T-V_T)\}$  are the terms on the right hand side of Equation~\ref{eq:gen_weighted_excess2}. The cross terms in the Jacobian matrix, $O(\sigma_{Wi,Wij})$ and $O(\sigma_{Wi,Wj})$ are proportional to the covariance of the uncertainties in the \WS\ photometry and the mean \WS\ colors. We ignore the first term, $O(\sigma_{Wi,Wij})$, because $\sigma_{Wij} \sim 0.1 \sigma_{Wi}$ and $W_{ij}$ is only a shallow function of $B_T-V_T$. We also ignore $O(\sigma_{Wi,Wj})$ because the errors on $Wi$ and $Wj$ are not correlated and hence $\sigma_{Wi,Wj}\sim 0$. Thus, Equation~\ref{eq:wtavgExcessUnc} reduces to

\begin{equation}\label{eq:wtavgExcessUnc_reduced}
      \sigma_{\overline{E[Wj]}}^2 \simeq \sum_{\alpha} \sigma_{\alpha}^2 \left(\frac{\partial \overline{E[Wj]}}{\partial \alpha}\right)^2,
\end{equation}

\noindent where $\alpha \in \{Wi, Wj\}$, after removing the photospheric uncertainties from the calculation. We define the significance of the weighted-color excess at $Wj$ in the same form as in Equation~\ref{eq:combined_significance}:

\begin{equation}\label{eq:combined_sig_appendix}
	    \Sigma_{\overline{E[Wj]}} = \frac{\overline{E[Wj]}}{\sigma_{\overline{E[Wj]}}}. 
\end{equation}

	We proceed with solving for the weights in equation~\ref{eq:gen_weighted_excess}. Using $j=4$ as an example, we can expand equation~\ref{eq:gen_weighted_excess} as 
	
\begin{eqnarray}\label{eq:W4_expanded_1}
\overline{E[W4]} & = & a_1 E[W1-W4] + a_2 E[W2-W4] + a_3 E[W3-W4] \\
				 & = & a_1 (W1-W4-W_{14}) + a_2(W2-W4-W_{24}) + a_3(W3-W4-W_{34}),
\end{eqnarray}

\noindent Inserting $a_3 = 1-a_1-a_2$ into Equation~\ref{eq:W4_expanded_1} produces

\begin{equation}\label{eq:w4_expanded_2}
\overline{E[W4]} =  a_1W1 - a_1W_{14} + a_2 W2 - a_2 W_{24} + W3 - W4 - W_{34} - a_1W3 + a_1 W_{34} - a_2 W3 + a_2 W_{34}. 
\end{equation}

\noindent The variance of $\overline{E[W4]}$ is calculated using Equation~\ref{eq:wtavgExcessUnc_reduced},

\begin{equation}\label{wtd_variance}
\sigma_{\overline{E[W4]}}^2 = a_1^2 \sigma_{W1}^2 + a_2^2 \sigma_{W2}^2 + (1-a_1-a_2)^2\sigma_{W3}^2  + \sigma_{W4}^2.
\end{equation}

\noindent Next we seek solutions for $a_1$ and $a_2$ that minimize the dependence of $\sigma_{\overline{E[W4]}}^2$ on these weights. Thus, by calculating

\begin{equation}\label{eq:mina1}
\left(\frac{\partial \sigma_{\overline{E[W4]}}^2}{\partial a_1}\right) = 0 = 2a_1\sigma_{W1}^2 - 2 \sigma_{W3}^2 + 2a_2\sigma_{W3}^2 + 2a_1\sigma_{W3}^2,
\end{equation}

\begin{equation}\label{eq:mina2}
\left(\frac{\partial \sigma_{\overline{E[W4]}}^2}{\partial a_2}\right) = 0 = 2a_2\sigma_{W2}^2 - 2 \sigma_{W3}^2 + 2a_2\sigma_{W3}^2 + 2a_1\sigma_{W3}^2
\end{equation}

We solve for $a_1$ and $a_2$

\begin{equation}\label{eq:a1solved}
a_1 = \frac{\sigma_{W3}^2\sigma_{W2}^2}{\sigma_{W2}^2\sigma_{W1}^2 + \sigma_{W2}^2\sigma_{W3}^2 + \sigma_{W3}^2\sigma_{W1}^2},
\end{equation}

\begin{equation}\label{eq:a2solved}
a_2 = \frac{\sigma_{W3}^2\sigma_{W1}^2}{\sigma_{W2}^2\sigma_{W1}^2 + \sigma_{W2}^2\sigma_{W3}^2 + \sigma_{W3}^2\sigma_{W1}^2}. 
\end{equation}

\noindent Now, using Equations~\ref{eq:a1solved} and \ref{eq:a2solved}, we recover $a_3$, 

\begin{equation}\label{eq:a3solved}
a_3 = \frac{\sigma_{W2}^2\sigma_{W1}^2}{\sigma_{W2}^2\sigma_{W1}^2 + \sigma_{W2}^2\sigma_{W3}^2 + \sigma_{W3}^2\sigma_{W1}^2}.
\end{equation}

To reduce the form of these weights, we multiply and divide each by $\sigma_{W1}^2\sigma_{W2}^2\sigma_{W3}^2$, to finally obtain the general form for each weight 

\begin{equation}\label{eq:ai}
a_i = \frac{1/\sigma_{Wi}^2}{\sum_{i=1}^{j-1}1/\sigma_{Wi}^2}. 
\end{equation}

\noindent  This is valid for either weighted $W3$ ($j=3$) or weighted $W4$ ($j=4$) excesses. We then set $A=\sum_{i=1}^{j-1}1/\sigma_{Wi}^2$, substitute equation \ref{eq:ai} into equation \ref{wtd_variance} to obtain a reduced expression for the variance of the excess ($\sigma_{\overline{E[W4]}}$), and then place that expression into Equation \ref{eq:combined_sig_appendix}. This gives us the final form for the significance of the weighted-color excess, which when generalized for $j=3$ or $j=4$ is

\begin{equation}\label{eq:wtd_significance_appendix2}
\Sigma_{\overline{E[Wj]}} = \frac{\frac{1}{A}\sum\limits_{i=1}^{j-1}\frac{E[Wi-Wj]}{\sigma_i^2}}{\sqrt{\sigma_j^2 + 1/A}}.  
\end{equation}

\noindent Equation~\ref{eq:wtd_significance_appendix2} is the same result for $\Sigma_{\overline{E[Wj]}}$ as presented in equation~\ref{eq:combined_significance}.

\clearpage

\bibliography{DebrisDisks}

\begin{thebibliography}{}
\expandafter\ifx\csname natexlab\endcsname\relax\def\natexlab#1{#1}\fi

\bibitem[{{Ardila} {et~al.}(2004){Ardila}, {Golimowski}, {Krist}, {Clampin},
  {Williams}, {Blakeslee}, {Ford}, {Hartig}, \& {Illingworth}}]{Ardila2004}
{Ardila}, D.~R., {Golimowski}, D.~A., {Krist}, J.~E., {et~al.} 2004, \apjl,
  617, L147

\bibitem[{{Ballering} {et~al.}(2013){Ballering}, {Rieke}, {Su}, \&
  {Montiel}}]{Ballering2013}
{Ballering}, N.~P., {Rieke}, G.~H., {Su}, K.~Y.~L., \& {Montiel}, E. 2013,
  \apj, 775, 55

\bibitem[{Balog {et~al.}(2014)Balog, Müller, Nielbock, Altieri, Klaas,
  Blommaert, Linz, Lutz, Moór, Billot, Sauvage, \& Okumura}]{Balog2013}
Balog, Z., Müller, T., Nielbock, M., {et~al.} 2014, Experimental Astronomy,
  37, 129

\bibitem[{{Beichman} {et~al.}(2006){Beichman}, {Bryden}, {Stapelfeldt},
  {Gautier}, {Grogan}, {Shao}, {Velusamy}, {Lawler}, {Blaylock}, {Rieke},
  {Lunine}, {Fischer}, {Marcy}, {Greaves}, {Wyatt}, {Holland}, \&
  {Dent}}]{Beichman2006}
{Beichman}, C.~A., {Bryden}, G., {Stapelfeldt}, K.~R., {et~al.} 2006, \apj,
  652, 1674

\bibitem[{{Brown} {et~al.}(2014){Brown}, {Jarrett}, \& {Cluver}}]{Brown2014}
{Brown}, M.~J.~I., {Jarrett}, T.~H., \& {Cluver}, M.~E. 2014, \pasa, 31, HASH

\bibitem[{{Bryden} {et~al.}(2006){Bryden}, {Beichman}, {Trilling}, {Rieke},
  {Holmes}, {Lawler}, {Stapelfeldt}, {Werner}, {Gautier}, {Blaylock}, {Gordon},
  {Stansberry}, \& {Su}}]{Bryden2006}
{Bryden}, G., {Beichman}, C.~A., {Trilling}, D.~E., {et~al.} 2006, \apj, 636,
  1098

\bibitem[{{Carpenter} {et~al.}(2009){Carpenter}, {Bouwman}, {Mamajek}, {Meyer},
  {Hillenbrand}, {Backman}, {Henning}, {Hines}, {Hollenbach}, {Kim},
  {Moro-Martin}, {Pascucci}, {Silverstone}, {Stauffer}, \&
  {Wolf}}]{Carpenter2009}
{Carpenter}, J.~M., {Bouwman}, J., {Mamajek}, E.~E., {et~al.} 2009, \apjs, 181,
  197

\bibitem[{{Chen} {et~al.}(2014){Chen}, {Mittal}, {Kuchner}, {Forrest}, {Lisse},
  {Manoj}, {Sargent}, \& {Watson}}]{Chen2014}
{Chen}, C.~H., {Mittal}, T., {Kuchner}, M., {et~al.} 2014, \apjs, 211, 25

\bibitem[{{Cotten} \& {Song}(2016)}]{Cotten2016}
{Cotten}, T.~H., \& {Song}, I. 2016, \apjs, 225, 15

\bibitem[{{Cruz-Saenz de Miera} {et~al.}(2014){Cruz-Saenz de Miera}, {Chavez},
  {Bertone}, \& {Vega}}]{Cruz-SaenzdeMiera2014}
{Cruz-Saenz de Miera}, F., {Chavez}, M., {Bertone}, E., \& {Vega}, O. 2014,
  \mnras, 437, 391

\bibitem[{{De Rosa} {et~al.}(2014){De Rosa}, {Patience}, {Wilson}, {Schneider},
  {Wiktorowicz}, {Vigan}, {Marois}, {Song}, {Macintosh}, {Graham}, {Doyon},
  {Bessell}, {Thomas}, \& {Lai}}]{DeRosa2014}
{De Rosa}, R.~J., {Patience}, J., {Wilson}, P.~A., {et~al.} 2014, \mnras, 437,
  1216

\bibitem[{{Dodson-Robinson} {et~al.}(2011){Dodson-Robinson}, {Beichman},
  {Carpenter}, \& {Bryden}}]{Dodson-Robinson2011}
{Dodson-Robinson}, S.~E., {Beichman}, C.~A., {Carpenter}, J.~M., \& {Bryden},
  G. 2011, \aj, 141, 11

\bibitem[{{Draper} {et~al.}(2016){Draper}, {Matthews}, {Kennedy}, {Wyatt},
  {Venn}, \& {Sibthorpe}}]{Draper2016}
{Draper}, Z.~H., {Matthews}, B.~C., {Kennedy}, G.~M., {et~al.} 2016, \mnras,
  456, 459

\bibitem[{{Eiroa} {et~al.}(2013){Eiroa}, {Marshall}, {Mora}, {Montesinos},
  {Absil}, {Augereau}, {Bayo}, {Bryden}, {Danchi}, {del Burgo}, {Ertel},
  {Fridlund}, {Heras}, {Krivov}, {Launhardt}, {Liseau}, {L{\"o}hne},
  {Maldonado}, {Pilbratt}, {Roberge}, {Rodmann}, {Sanz-Forcada}, {Solano},
  {Stapelfeldt}, {Th{\'e}bault}, {Wolf}, {Ardila}, {Ar{\'e}valo}, {Beichmann},
  {Faramaz}, {Gonz{\'a}lez-Garc{\'{\i}}a}, {Guti{\'e}rrez}, {Lebreton},
  {Mart{\'{\i}}nez-Arn{\'a}iz}, {Meeus}, {Montes}, {Olofsson}, {Su}, {White},
  {Barrado}, {Fukagawa}, {Gr{\"u}n}, {Kamp}, {Lorente}, {Morbidelli},
  {M{\"u}ller}, {Mutschke}, {Nakagawa}, {Ribas}, \& {Walker}}]{Eiroa2013}
{Eiroa}, C., {Marshall}, J.~P., {Mora}, A., {et~al.} 2013, \aap, 555, A11

\bibitem[{{Fujiwara} {et~al.}(2013){Fujiwara}, {Ishihara}, {Onaka}, {Takita},
  {Kataza}, {Yamashita}, {Fukagawa}, {Ootsubo}, {Hirao}, {Enya}, {Marshall},
  {White}, {Nakagawa}, \& {Murakami}}]{Fujiwara2013}
{Fujiwara}, H., {Ishihara}, D., {Onaka}, T., {et~al.} 2013, \aap, 550, A45

\bibitem[{{Gray} {et~al.}(2006){Gray}, {Corbally}, {Garrison}, {McFadden},
  {Bubar}, {McGahee}, {O'Donoghue}, \& {Knox}}]{Gray2006}
{Gray}, R.~O., {Corbally}, C.~J., {Garrison}, R.~F., {et~al.} 2006, \aj, 132,
  161

\bibitem[{{Houck} {et~al.}(2004){Houck}, {Roellig}, {van Cleve}, {Forrest},
  {Herter}, {Lawrence}, {Matthews}, {Reitsema}, {Soifer}, {Watson}, {Weedman},
  {Huisjen}, {Troeltzsch}, {Barry}, {Bernard-Salas}, {Blacken}, {Brandl},
  {Charmandaris}, {Devost}, {Gull}, {Hall}, {Henderson}, {Higdon}, {Pirger},
  {Schoenwald}, {Sloan}, {Uchida}, {Appleton}, {Armus}, {Burgdorf},
  {Fajardo-Acosta}, {Grillmair}, {Ingalls}, {Morris}, \& {Teplitz}}]{Houck2004}
{Houck}, J.~R., {Roellig}, T.~L., {van Cleve}, J., {et~al.} 2004, \apjs, 154,
  18

\bibitem[{Hunter(2007)}]{Hunter2007}
Hunter, J.~D. 2007, Computing In Science \& Engineering, 9, 90

\bibitem[{{Lallement} {et~al.}(2003){Lallement}, {Welsh}, {Vergely}, {Crifo},
  \& {Sfeir}}]{Lallement2003}
{Lallement}, R., {Welsh}, B.~Y., {Vergely}, J.~L., {Crifo}, F., \& {Sfeir}, D.
  2003, \aap, 411, 447

\bibitem[{{Lang}(2014)}]{Lang2014}
{Lang}, D. 2014, \aj, 147, 108

\bibitem[{{Lawler} {et~al.}(2009){Lawler}, {Beichman}, {Bryden}, {Ciardi},
  {Tanner}, {Su}, {Stapelfeldt}, {Lisse}, \& {Harker}}]{Lawler2009}
{Lawler}, S.~M., {Beichman}, C.~A., {Bryden}, G., {et~al.} 2009, \apj, 705, 89

\bibitem[{Mahalanobis(1936)}]{Mahalanobis1936}
Mahalanobis, P.~C. 1936, Proceedings of the National Institute of Sciences
  (Calcutta), 2, 49

\bibitem[{{McDonald} {et~al.}(2012){McDonald}, {Zijlstra}, \&
  {Boyer}}]{McDonald2012}
{McDonald}, I., {Zijlstra}, A.~A., \& {Boyer}, M.~L. 2012, \mnras, 427, 343

\bibitem[{{Mizusawa} {et~al.}(2012){Mizusawa}, {Rebull}, {Stauffer}, {Bryden},
  {Meyer}, \& {Song}}]{Mizusawa2012}
{Mizusawa}, T.~F., {Rebull}, L.~M., {Stauffer}, J.~R., {et~al.} 2012, \aj, 144,
  135

\bibitem[{{Mo{\'o}r} {et~al.}(2006){Mo{\'o}r}, {{\'A}brah{\'a}m}, {Derekas},
  {Kiss}, {Kiss}, {Apai}, {Grady}, \& {Henning}}]{Moor2006}
{Mo{\'o}r}, A., {{\'A}brah{\'a}m}, P., {Derekas}, A., {et~al.} 2006, \apj, 644,
  525

\bibitem[{{Patel} {et~al.}(2014{\natexlab{a}}){Patel}, {Metchev}, \&
  {Heinze}}]{Patel2014}
{Patel}, R.~I., {Metchev}, S.~A., \& {Heinze}, A. 2014{\natexlab{a}}, \apjs,
  212, 10

\bibitem[{{Patel} {et~al.}(2014{\natexlab{b}}){Patel}, {Metchev}, \&
  {Heinze}}]{Patel2014b}
---. 2014{\natexlab{b}}, \apjs, 214, 14

\bibitem[{{Patel} {et~al.}(2015){Patel}, {Metchev}, \& {Heinze}}]{Patel2015}
---. 2015, \apjs, 220, 21

\bibitem[{{Pecaut} \& {Mamajek}(2013)}]{Pecaut2013}
{Pecaut}, M.~J., \& {Mamajek}, E.~E. 2013, \apjs, 208, 9

\bibitem[{{Perryman} {et~al.}(1997){Perryman}, {Lindegren}, {Kovalevsky},
  {Hoeg}, {Bastian}, {Bernacca}, {Cr{\'e}z{\'e}}, {Donati}, {Grenon},
  {Grewing}, {van Leeuwen}, {van der Marel}, {Mignard}, {Murray}, {Le Poole},
  {Schrijver}, {Turon}, {Arenou}, {Froeschl{\'e}}, \&
  {Petersen}}]{Perryman1997}
{Perryman}, M.~A.~C., {Lindegren}, L., {Kovalevsky}, J., {et~al.} 1997, \aap,
  323, L49

\bibitem[{{Plavchan} {et~al.}(2009){Plavchan}, {Werner}, {Chen}, {Stapelfeldt},
  {Su}, {Stauffer}, \& {Song}}]{Plavchan2009}
{Plavchan}, P., {Werner}, M.~W., {Chen}, C.~H., {et~al.} 2009, \apj, 698, 1068

\bibitem[{{Rebull} {et~al.}(2008){Rebull}, {Stapelfeldt}, {Werner}, {Mannings},
  {Chen}, {Stauffer}, {Smith}, {Song}, {Hines}, \& {Low}}]{Rebull2008}
{Rebull}, L.~M., {Stapelfeldt}, K.~R., {Werner}, M.~W., {et~al.} 2008, \apj,
  681, 1484

\bibitem[{{Rhee} {et~al.}(2007){Rhee}, {Song}, {Zuckerman}, \&
  {McElwain}}]{Rhee2007}
{Rhee}, J.~H., {Song}, I., {Zuckerman}, B., \& {McElwain}, M. 2007, \apj, 660,
  1556

\bibitem[{{Ricci} {et~al.}(2015){Ricci}, {Carpenter}, {Fu}, {Hughes}, {Corder},
  \& {Isella}}]{Ricci2015}
{Ricci}, L., {Carpenter}, J.~M., {Fu}, B., {et~al.} 2015, \apj, 798, 124

\bibitem[{{Riviere-Marichalar} {et~al.}(2014){Riviere-Marichalar}, {Barrado},
  {Montesinos}, {Duch{\^e}ne}, {Bouy}, {Pinte}, {Menard}, {Donaldson}, {Eiroa},
  {Krivov}, {Kamp}, {Mendigut{\'{\i}}a}, {Dent}, \&
  {Lillo-Box}}]{Riviere-Marichalar2014}
{Riviere-Marichalar}, P., {Barrado}, D., {Montesinos}, B., {et~al.} 2014, \aap,
  565, A68

\bibitem[{{Rizzuto} {et~al.}(2012){Rizzuto}, {Ireland}, \&
  {Zucker}}]{Rizzuto2012}
{Rizzuto}, A.~C., {Ireland}, M.~J., \& {Zucker}, D.~B. 2012, \mnras, 421, L97

\bibitem[{{Robitaille} \& {Bressert}(2012)}]{Robitaille2012}
{Robitaille}, T., \& {Bressert}, E. 2012, {APLpy: Astronomical Plotting Library
  in Python}, Astrophysics Source Code Library, , , ascl:1208.017

\bibitem[{Rousseeuw \& Driessen(1999)}]{Rousseeuw1999}
Rousseeuw, P.~J., \& Driessen, K.~V. 1999, Technometrics, 41, 212

\bibitem[{{Stauffer} {et~al.}(2010){Stauffer}, {Rebull}, {James},
  {Noriega-Crespo}, {Strom}, {Wolk}, {Carpenter}, {Barrado y Navascues},
  {Micela}, {Backman}, \& {Cargile}}]{Stauffer2010}
{Stauffer}, J.~R., {Rebull}, L.~M., {James}, D., {et~al.} 2010, \apj, 719, 1859

\bibitem[{{Su} {et~al.}(2006){Su}, {Rieke}, {Stansberry}, {Bryden},
  {Stapelfeldt}, {Trilling}, {Muzerolle}, {Beichman}, {Moro-Martin}, {Hines},
  \& {Werner}}]{Su2006}
{Su}, K.~Y.~L., {Rieke}, G.~H., {Stansberry}, J.~A., {et~al.} 2006, \apj, 653,
  675

\bibitem[{{Theissen} \& {West}(2014)}]{Theissen2014}
{Theissen}, C.~A., \& {West}, A.~A. 2014, \apj, 794, 146

\bibitem[{{Trilling} {et~al.}(2008){Trilling}, {Bryden}, {Beichman}, {Rieke},
  {Su}, {Stansberry}, {Blaylock}, {Stapelfeldt}, {Beeman}, \&
  {Haller}}]{Trilling2008}
{Trilling}, D.~E., {Bryden}, G., {Beichman}, C.~A., {et~al.} 2008, \apj, 674,
  1086

\bibitem[{Urban {et~al.}(2012)Urban, Rieke, Su, \& Trilling}]{Urban2012}
Urban, L.~E., Rieke, G., Su, K., \& Trilling, D.~E. 2012, The Astrophysical
  Journal, 750, 98

\bibitem[{{van Leeuwen}(2007)}]{VanLeeuwen2007}
{van Leeuwen}, F. 2007, \aap, 474, 653

\bibitem[{{Vican} \& {Schneider}(2014)}]{Vican2014}
{Vican}, L., \& {Schneider}, A. 2014, \apj, 780, 154

\bibitem[{{Wahhaj} {et~al.}(2015){Wahhaj}, {Cieza}, {Mawet}, {Yang}, {Canovas},
  {De Boer}, {Casassus}, {Menard}, {Schreiber}, {Liu}, {Biller}, {Nielsen}, \&
  {Hayward}}]{Wahhaj2015}
{Wahhaj}, Z., {Cieza}, L.~A., {Mawet}, D., {et~al.} 2015, ArXiv e-prints,
  arXiv:1502.03092

\bibitem[{{Wright} {et~al.}(2010){Wright}, {Eisenhardt}, {Mainzer}, {Ressler},
  {Cutri}, {Jarrett}, {Kirkpatrick}, {Padgett}, {McMillan}, {Skrutskie},
  {Stanford}, {Cohen}, {Walker}, {Mather}, {Leisawitz}, {Gautier}, {McLean},
  {Benford}, {Lonsdale}, {Blain}, {Mendez}, {Irace}, {Duval}, {Liu}, {Royer},
  {Heinrichsen}, {Howard}, {Shannon}, {Kendall}, {Walsh}, {Larsen}, {Cardon},
  {Schick}, {Schwalm}, {Abid}, {Fabinsky}, {Naes}, \& {Tsai}}]{Wright2010}
{Wright}, E.~L., {Eisenhardt}, P.~R.~M., {Mainzer}, A.~K., {et~al.} 2010, \aj,
  140, 1868

\bibitem[{{Wu} {et~al.}(2013){Wu}, {Wu}, {Lam}, {Yang}, {Wen}, {Li}, {Zhang},
  \& {Gao}}]{Wu2013}
{Wu}, C.-J., {Wu}, H., {Lam}, M.-I., {et~al.} 2013, \apjs, 208, 29

\bibitem[{{Zuckerman}(2001)}]{Zuckerman2001}
{Zuckerman}, B. 2001, \araa, 39, 549

\bibitem[{{Zuckerman} \& {Song}(2004)}]{Zuckerman2004}
{Zuckerman}, B., \& {Song}, I. 2004, \apj, 603, 738

\end{thebibliography}


\end{document}